\documentclass[aps, nofootinbib,floatfix, amsmath, amssymb, runinaddress, tightenlines, 11pt]{revtex4-2}
\pdfoutput=1
\usepackage{caption}
\usepackage{subcaption}
\usepackage{graphicx}
\usepackage{hyperref}
\usepackage{amsmath}
\usepackage{mathtools}
\usepackage{upgreek}
\usepackage{tablefootnote}
\usepackage{soul}
\usepackage{color}
\usepackage{verbatim}
\usepackage{float}
\usepackage{bm}
\usepackage{placeins}
\usepackage{enumitem}
\usepackage[export]{adjustbox}

\DeclareUnicodeCharacter{2212}{-}

\usepackage[usenames,dvipsnames,svgnames,table]{xcolor}

\usepackage{simplewick}
\usepackage{simpler-wick}
\usepackage{amssymb}
\usepackage{epstopdf}
\graphicspath{ {./Figures/} }
\usepackage{dcolumn}
\usepackage{bm}
\usepackage[normalem]{ulem}
\usepackage{multirow}
\usepackage{booktabs}
\usepackage{contour}
\usetikzlibrary{positioning}
\usetikzlibrary{shapes.geometric}

\newcommand{\bsl}{\boldsymbol{\ell}}

\def\q{{\boldsymbol{q}}}

\def\ktwo{\boldsymbol{k}_2}
\def\kthree{\boldsymbol{k}_3}

\def\qone{\boldsymbol{q}_1}
\def\qtwo{\boldsymbol{q}_2}

\newcommand{\bk}[1]{\boldsymbol{k}_{#1}}

\newcommand{\fnl}[1]{f_\text{NL}^{\text{#1}}}

\def\zk{(z,\boldsymbol{k})}

\def\beq{\begin{equation}}
\def\eeq{\end{equation}}

\def\db{\delta_b}

\def\dTcm{\delta T_{21}}

\def\camb{\texttt{CAMB}}

\begin{document}
\author{Giorgio Orlando$^{1}$}\email{g.orlando@rug.nl}
\author{Thomas Fl\"{o}ss$^{1,2}$}\email{t.s.floss@rug.nl}
\author{P. Daniel Meerburg$^{1}$}\email{p.d.meerburg@rug.nl}
\author{Joseph Silk$^{3,4,5}$}\email{silk@iap.fr}
\affiliation{\\ 1) Van Swinderen Institute for Particle Physics and Gravity, University of
Groningen, Nijenborgh 4, 9747 AG Groningen, The Netherlands.}
\affiliation{\\ 2) Kapteyn Astronomical Institute, University of Groningen, P.O.Box 800, 9700 AV Groningen, The Netherlands}
\affiliation{\\ 3) Institut d'Astrophysique de Paris, UMR 7095 CNRS, Sorbonne University, 75014 Paris, France}
\affiliation{\\ 4) Department of Physics and Astronomy, Johns Hopkins University,
Baltimore, MD 21218, USA}
\affiliation{\\ 5) Beecroft Institute of Particle Astrophysics and Cosmology, Department of Physics, University of Oxford, Oxford OX1 3RH, UK}
\date{\today}
\title{Local non-Gaussianities from cross-correlations between the CMB and 21-cm}

\begin{abstract}

The 21-cm brightness temperature fluctuation from the Dark Ages ($z \simeq 30-100$) will allow us to probe the inflationary epoch on very small scales ($>0.1 \, \mbox{Mpc}^{-1}$), inaccessible to cosmic microwave background experiments. Combined with the possibility to collect information from different redshift slices, the 21-cm bispectrum has the potential to significantly improve constraints on primordial non-Gaussianity. 
However, recent work has shown secondary effects source off-diagonal terms in the covariance matrix which can significantly affect forecasted constraints, especially in signals that peak in the squeezed configuration, such as the local bispectrum. 
In this paper we propose the three-point $\langle 21-21-\rm CMB \rangle$ bispectrum cross-correlation as a new independent observational channel sensitive to local primordial non-Gaussianity. 
We find that, contrary to the 21-cm bispectrum, secondary contributions are subdominant to the primordial signal for values $f_{\rm NL}^{\rm loc} \sim 1$, resulting in negligible effects from off-diagonal terms in the covariance matrix. 
We forecast that CMB $T$ and $E$ modes cross-correlated with an ideal cosmic variance-limited 21-cm experiment with a $0.1$ MHz frequency and $0.1$ arc-minute angular resolution could reach $f_{\rm NL}^{\rm loc} \sim 6 \times 10^{-3}$. This forecast suggests cross-correlation between CMB and 21-cm experiments could provide a viable alternative to 21-cm auto-spectra in reaching unprecedented constraints on primordial local non-Gaussianities.
\end{abstract}
 
\maketitle
\newpage
\tableofcontents
\newpage


\section{Introduction}

The inflationary model has become the dominant paradigm for describing the early universe, and it accurately predicts the nearly Gaussian statistics of the Cosmic Microwave Background (CMB) anisotropies \cite{Planck:2018vyg,Planck:2019kim}. However, small deviations from  Gaussianity, known as primordial non-Gaussianity (pnG), can be used to further constrain the vast space of inflationary theories that go beyond a simple single-field slow-roll scenario. We currently lack a detection of a ‘gravitational floor’ which sets the minimum amount of pnG that should be present in the initial conditions sourced by inflation \cite{Maldacena:2002vr}. Our limitation in the attempt to measure pnG is the smallness of the signal versus the finite range of scales that CMB experiments are capable of probing ($k \lesssim 0.1 \, \mbox{Mpc}^{-1}$).

In recent years, the bispectrum of the 21-cm brightness temperature fluctuations emitted during the cosmic Dark Ages has been proposed as an ultimate probe of pnG (see e.g. \cite{Cooray:2006km,Pillepich:2006fj,Meerburg:2016zdz,Munoz:2015eqa,Silk:2020bsr,Floss:2022grj}). 21-cm fluctuations trace the matter density field which, in turn, traces the primordial fluctuations seeded by inflation. The 21-cm field allows us to probe very small scales up to $k \sim 10 \, \mbox{Mpc}^{-1}$ for arc-minute angular resolution experiments (like those proposed for the far-side of the moon, see e.g. \cite{Cole:2019zhu}). This would allow us to probe non-Gaussian amplitudes of the order $f_{\rm NL}\sim 10^{-1}$. Combining complementary information from different redshift slices, we can further improve this constraint to $f_{\rm NL}\sim 10^{-2}$. The first studies that have been carried out (e.g. \cite{Pillepich:2006fj,Munoz:2015eqa}) assume a cosmic-variance limited detection of the 21-cm field and a diagonal covariance matrix for the 21-cm bispectrum. Recent papers (see e.g. \cite{Biagetti:2021tua, Floss:2022wkq}) suggest that off-diagonal terms on the covariance matrix have a significant impact on pnG constraints, particularly on the local shape, resulting in the degradation of the $f_{\rm NL}^{\rm loc}$ minimum detectable value of more than one order of magnitude per redshift slice. 

In this paper we propose a new observable, which could perhaps limit the effect of nG covariance, but maintain the large signal by probing very small scales. We study the three-point cross-correlations between two 21-cm anisotropy fields emitted at a given redshift slice $z$ and a CMB anisotropy field (both temperature and polarization). As CMB anisotropies probe primordial scales that are much larger than the 21-cm tracer field, this observable is naturally sensitive to shape functions that peak in squeezed triangular configurations, such as the local shape. We compute the primordial effect introduced by local pnG and the secondary effects generated by non-linear evolution. We find that, contrary to the 21-cm auto bispectrum, as long as CMB anisotropies are considered on very large scales ($\ell_{\rm CMB} \lesssim 10$ or $k< 10^{-3} \, \mbox{Mpc}^{-1}$) the secondary contribution is generally subdominant to the primordial signal for $f_{\rm NL}^{\rm loc} = 1$. The intuitive reason is that secondary contributions are proportional to the baryon power spectrum evaluated on the large mode $k_l$ of squeezed configurations. In the cross-correlations this mode is carried by CMB anisotropies ($k_l< 10^{-3} \, \mbox{Mpc}^{-1}$), providing a very small large-scale baryon power spectrum (see bottom panel of Fig. \ref{fig:est}). As a consequence we find that secondary contributions, if modeled with a reasonable percent accuracy level, can in principle be removed from the data, in a similar way to the analysis of secondary contributions in the CMB auto bispectrum. We perform a Fisher forecast assuming a cosmic variance limited detection of the 21-cm field up to $k \sim 10 \, \mbox{Mpc}^{-1}$ ($\ell_{\rm 21cm} \simeq 10^5$) and a CMB field detection up to $k \sim 10^{-3} \, \mbox{Mpc}^{-1}$ ($\ell_{\rm CMB} = 10$), which contribute the most to the Fisher information. We find that using the $\langle 21-21-\rm T \rangle$ cross-correlation, $f_{\rm NL}^{\rm loc} \sim 2 \times 10^{-1}$ can be achieved for a single 21-cm redshift slice. A slight improvement of a factor $2$ can be reached by implementing the information of the polarization field $E$. Finally, considering tomography of the 21-cm field of the entire Dark Ages, $f_{\rm NL}^{\rm loc} \sim 6 \times 10^{-3}$ could be achieved with a frequency resolution $\Delta \nu = 0.1$ MHz. As secondary contributions are small, we estimate that off-diagonal terms in the covariance matrix should not significantly alter this forecast. Also, this measurement refers to squeezed triangular configurations ($k_l < 10^{-3} \, \mbox{Mpc}^{-1}$, $k_s> 0.1 \, \mbox{Mpc}^{-1}$) that are complementary to those obtained from CMB ($k_l, k_s \lesssim 0.1 \, \mbox{Mpc}^{-1}$) and 21-cm ($k_l, k_s \gtrsim 0.1 \, \mbox{Mpc}^{-1}$) auto-correlations.  

In the present analysis we neglect velocity-terms in the 21-cm field, assuming the latter to be independent of the line-of-sight direction. This is a good approximation given the 21-cm scales considered. We thus expect our results to be qualitatively valid in general. We do not address foreground contamination, leaving this for future analysis. According to our results, provided that a cosmic-variance limited detection of the 21-cm anisotropies can be made, the final forecasts on $f_{\rm NL}^{\rm loc}$ are not dependent on the specific 21-cm amplitude, but only on the angular and frequency resolution of a given 21-cm experiment. 

The paper is organized as follows. In Sec. \ref{sec:basics} we provide some background on inflation, defining the local bispectrum. In Sec. \ref{sec:analanis} we provide our conventions for the analytical expressions of the CMB and 21-cm anisotropy fields. In Sec. \ref{sec:2121X} we compute primordial and secondary contributions to the $\langle 21-21-\rm CMB \rangle$ bispectrum. In Sec. \ref{sec:fisher} we perform the Fisher forecasts. In Sec. \ref{sec:concl} we discuss the scientific results obtained and present our conclusions. The Appendixes contain some technical details of our calculations.  

\section{Basics} \label{sec:basics}

In this section we will introduce our convention for describing primordial (scalar) perturbations from inflation. First, we define the Fourier transform of scalar perturbations as
\begin{equation}
\zeta(\mathbf{x}) = \int \frac{d^3 k}{(2 \pi)^3} \,  e^{i \mathbf{k} \cdot \mathbf{x}} \, \zeta_{\mathbf{k}} \, .
\end{equation}
We define the primordial scalar power spectrum as
\begin{equation}
\langle \zeta_{\mathbf{k_1}} \zeta_{\mathbf{k_2}}\rangle = (2 \pi)^3 \delta^{(3)}(\mathbf{k_1} + \mathbf{k_2}) \, P_\zeta(k_1) \, .
\end{equation}
The power spectrum of primordial scalar perturbations from inflation can be expressed as
\begin{equation} \label{eq:power_inflation}
P_{\zeta}(k) = \frac{2 \pi^2}{k^3}  \mathcal A_s(k)  \, ,
\end{equation}
where $\mathcal A_s(k)$ is the dimensionless amplitude as measured by CMB experiments. Finally, we define the primordial scalar bispectrum
\begin{equation} \label{eq:def_bispectrum_scalar}
\langle \zeta_{\mathbf{k_1}} \zeta_{\mathbf{k_2}} \zeta_{\mathbf{k_3}}\rangle = (2 \pi)^3 \delta^{(3)}(\mathbf{k_1}+ \mathbf{k_2}+\mathbf{k_3}) \, B_{\zeta\zeta\zeta}(k_1, k_2, k_3) \, ,
\end{equation}
where we assume invariance under translations and rotation. 

Given the specific momentum-dependence of the bispectrum we can have several shapes ($B_{\zeta\zeta\zeta}$), each associated with specific physical mechanisms arising during inflation (see e.g. \cite{Baumann:2018muz} for a review). In this work we are interested in probing the so-called local configuration, which is given by 
\begin{equation}
    \label{eq:BLocal}
    B_{\zeta\zeta\zeta}^{\text{loc}}(k_1,k_2,k_3) = \frac{6}{5} \, \fnl{loc} \, \Big( P_\zeta(k_1)P_\zeta(k_2) + P_\zeta(k_2)P_\zeta(k_3) + P_\zeta(k_1)P_\zeta(k_3)\Big) \, .
\end{equation}
This shape peaks in the squeezed triangle configuration (e.g. $k_1 \ll k_2 \approx k_3$). 
A sizeable local bispectrum ($\fnl{loc} \gtrsim 1$) naturally arises in multi-field models of inflation (see e.g. \cite{Byrnes:2010em} for a review), where extra light fields modulate the inflationary dynamics. As the observable we consider in this paper is sensitive to squeezed triangular configurations, our focus is on observational prospects of this specific shape of non-Gaussianity.

\begin{table}[h!]
    \begingroup
    \setlength{\tabcolsep}{8pt} 
    \renewcommand{\arraystretch}{1.5} 
    \centering
    \begin{tabular}{c  c  c}
     \toprule
     &  \textbf{Parameters input in \texttt{CAMB}} &\\
     \midrule
        $H_0= 67.32 \, \mbox{km}/\mbox{s} \, \mbox{Mpc}^{-1}$ &  $\Omega_\mathrm{b} h^2 = 0.022383$ & $\Omega_\mathrm{c} h^2 = 0.12011$ \\
        $\Omega_\mathrm{k} = 0$ &  $\Omega_\mathrm{c} h^2 = 0.12011$  & $\tau = 0.0543$\\
        \bottomrule
    \end{tabular}
    \endgroup
    \caption{Best-fit \textit{Planck} parameters obtained combining $TT$, $TE$, $EE$+low$E$+lensing (see the \textit{Plik} best-fit of Tab.~1 of Ref.~\cite{Planck:2018vyg}).}
    \label{tab:cosm_par}
\end{table}
\FloatBarrier

\section{Analytical expressions for the anisotropies} \label{sec:analanis}

\subsection{CMB}

The coefficients of the unpolarized $X=T, E$-mode polarization anisotropies sourced by the scalar curvature perturbation from inflation ($\zeta$), can be connected to the scalar perturbations Fourier transform via \cite{Shiraishi:2010sm,Shiraishi:2010kd}
\begin{align} 
a_{\ell m}^{(s) X} &=
4\pi \, i^{\ell} \, \int \frac{d^3 p}{(2\pi)^{3}}
{\cal T}_{\ell(s)}^{X}(p) \, Y_{\ell m}^*(\hat{p}) \,\zeta_{\mathbf{p}}  \, . \label{eq:a_CMB_scalar} 
\end{align}
Here ${\cal T}_{\ell (s)}^{X}(p)$ is the scalar transfer function for the field $X \in (T, E)$. In this work we evaluate this transfer function using the publicly available Boltzmann solver \camb{} \cite{camb_notes}, whose main cosmological parameters are summarized in Tab. \ref{tab:cosm_par} 

The CMB power spectrum can be calculated as
\begin{align}  \label{eq:Xspectrum}
C^{\rm X}_{\ell_1 \ell_2} =& \frac{1}{2 \ell_1+1} \sum_{m_1, \, m_2} \, \langle a_{\ell_1 m_1}^{(s) X} a_{\ell_2 m_2}^{(s) X *} \rangle \nonumber \\
=& \frac{1}{2 \ell_1+1} \sum_{m_1, \, m_2} \,  (4\pi)^2 \, i^{\ell_1-\ell_2} \int \frac{d p \, p^2}{(2\pi)^{3}} \, \left({\cal T}_{\ell_1(s)}^{X}(p) \, {\cal T}_{\ell_2(s)}^{X}(p)\right) \, P_\zeta(p)  \, \int d \hat p   \, Y_{\ell_1 m_1}^*(\hat{p})  \, Y_{\ell_2 m_2}(\hat{p})   \nonumber \\
=& \delta_{\ell_1 \ell_2} \, 4 \pi \, \int \frac{d p}{p} \, \left({\cal T}_{\ell_1(s)}^{X}(p)\right)^2\,\mathcal A_s(p) \, .
\end{align}

\subsection{21-cm brightness temperature}

Similarly, we can define the 21-cm full-sky anisotropies in terms of the 21-cm brightness temperature Fourier transform $\dTcm(r'(z),\bk{})$ as \cite{Meerburg:2013dua}
\begin{align} 
a_{\ell m}^{21} &=
4\pi \, i^{\ell} \, \int \frac{d^3 k}{(2\pi)^{3}}
 \, Y_{\ell m}^*(\hat{k}) \,\alpha^{21}_\ell(\mathbf{k},z)  \, , \label{eq:21_cm_anisotropies}
\end{align}
where
\begin{align}
\alpha^{21}_\ell(\mathbf{k},z) =  \int_0^\infty d r' \, W_{r(z)}(r') \,  \dTcm(r',\bk{}) \, j_\ell(k r') \, ,
\end{align}
with $r' \equiv r'(z)$ denoting the comoving distance at a given redshift, $j_\ell(x)$ is a spherical Bessel function and $W_{r(z)}(r')$ is the frequency-dependent instrumental response function (from here on window function). Here we adopt a Gaussian window function
\begin{equation} \label{eq:window}
W_{r(z)}(r') = \frac{1}{\delta r \sqrt{2 \pi}} \, \exp\left[-\frac{1}{2} \left( \frac{r' - r}{\delta r}\right)^2\right] \, ,
\end{equation}
with width \cite{Meerburg:2013dua}
\begin{equation} \label{eq:deltar}
\delta r \simeq \left( \frac{\Delta \nu}{0.1 \, \mbox{MHz}}\right) \, \left( \frac{1+z}{10} \right)^{1/2} \, \left( \frac{\Omega_m h^2}{0.15}\right)^{-1/2} \, \mbox{Mpc}  \, ,
\end{equation}
where $\Delta \nu$ is the frequency bandwidth of a hypothetical survey instrument. \\

The Fourier transform  of the 21-cm brightness temperature $\dTcm(r'(z),\bk{})$ is modeled in several references, e.g. \cite{Lewis:2007kz, Pillepich:2006fj, Munoz:2015eqa, Floss:2022grj} and can be expressed in a power series of the baryon density $\db$ and velocity divergence $\theta_b$ fluctuations
\begin{equation}
    \db(z,\bk{}) =\sum_{n}\db^{(n)}(z,\bk{}) \, , \qquad \mbox{and}  \qquad \theta_b(z,\bk{}) = \sum_{n} \theta_b^{(n)}(z,\bk{}) \, ,
\end{equation}
where $\db^{(n)}(z,\bk{})$, $\theta_b^{(n)}(z,\bk{})$ denote n-th order terms in the perturbative expansion. Neglecting velocity-terms, which provide subdominant contributions to the small scales considered in this work, to second order
\begin{align}
	\dTcm\zk =&\;\dTcm^{(1)}\zk+\dTcm^{(2)}\zk \, ,
\end{align}
with $\dTcm^{(i)}\zk$ 
\begin{align}\label{eq:dT21expansion}
	\dTcm^{(1)}\zk=&\;\alpha_1(z)  \, \db^{(1)}(\bk{}) \, ,\nonumber\\
	\dTcm^{(2)}\zk=&\; \alpha_2(z) \, \db^{(2)}(\bk{})+ \alpha_3(z) \, \int \frac{d^3 q}{(2 \pi)^3}  \, \db^{(1)}(\q) \, \db^{(1)}(\bk{}-\q) \, .
\end{align}
Here the redshift dependence in $\db^{(n)}$ is re-absorbed in the $\alpha_i(z)$ coefficients\footnote{These correspond to coefficients $\alpha$, $\beta$ and $\gamma$ of Ref. \cite{Munoz:2015eqa} as: $\alpha_1 =\alpha$, $\, \alpha_2 =\beta$,  $\, \alpha_3 =\gamma$.}. 
The first order baryon over-density can be connected to the primordial perturbation from inflation $\zeta$ via
\begin{equation}
\delta^{(1)}_{b}(\bk{}, z) = \mathcal{M}_b(k, z) \,  \zeta(\bk{}) \, , 
\end{equation}
where $\mathcal{M}_b$ is the linear transfer function of baryon fluctuations (which can be obtained through e.g. \camb{}). Using standard perturbation theory (SPT) the second order baryon density perturbations $\db^{(2)}$ can be expressed in terms of the first order $\db^{(1)}$ as (see e.g. \cite{Bernardeau:2010ac})
\begin{equation}
	\delta^{(2)}_{b}(\bk{})\equiv  \int \frac{d^3 q_1}{(2 \pi)^3}  \int \frac{d^3 q_2}{(2 \pi)^3} \, (2\pi)^3 \, \delta_{\rm D}^{(3)}(\bk{}-\qone-\qtwo)\; F_2^{(s)}(\qone,\qtwo)\;\delta^{(1)}_{b}(\qone) \, \delta^{(1)}_{b}(\qtwo) \, ,
\end{equation}
where we introduced the kernel
\begin{equation} \label{eq:kernel}
 F_2^{(s)}(\qone,\qtwo) = c_1 + c_2 \, (\hat q_1 \cdot \hat q_2) \left(\frac{q_1}{q_2} + \frac{q_2}{q_1} \right) + c_3 \, (\hat q_1 \cdot \hat q_2)^2 \, , 
\end{equation}
with $c_1 = 5/7, \, c_2 = 1/2, \, c_3 = 2/7$ for a cold dark matter (CDM)-universe. \\

\begin{figure}
        \centering
        \includegraphics[width=\textwidth]{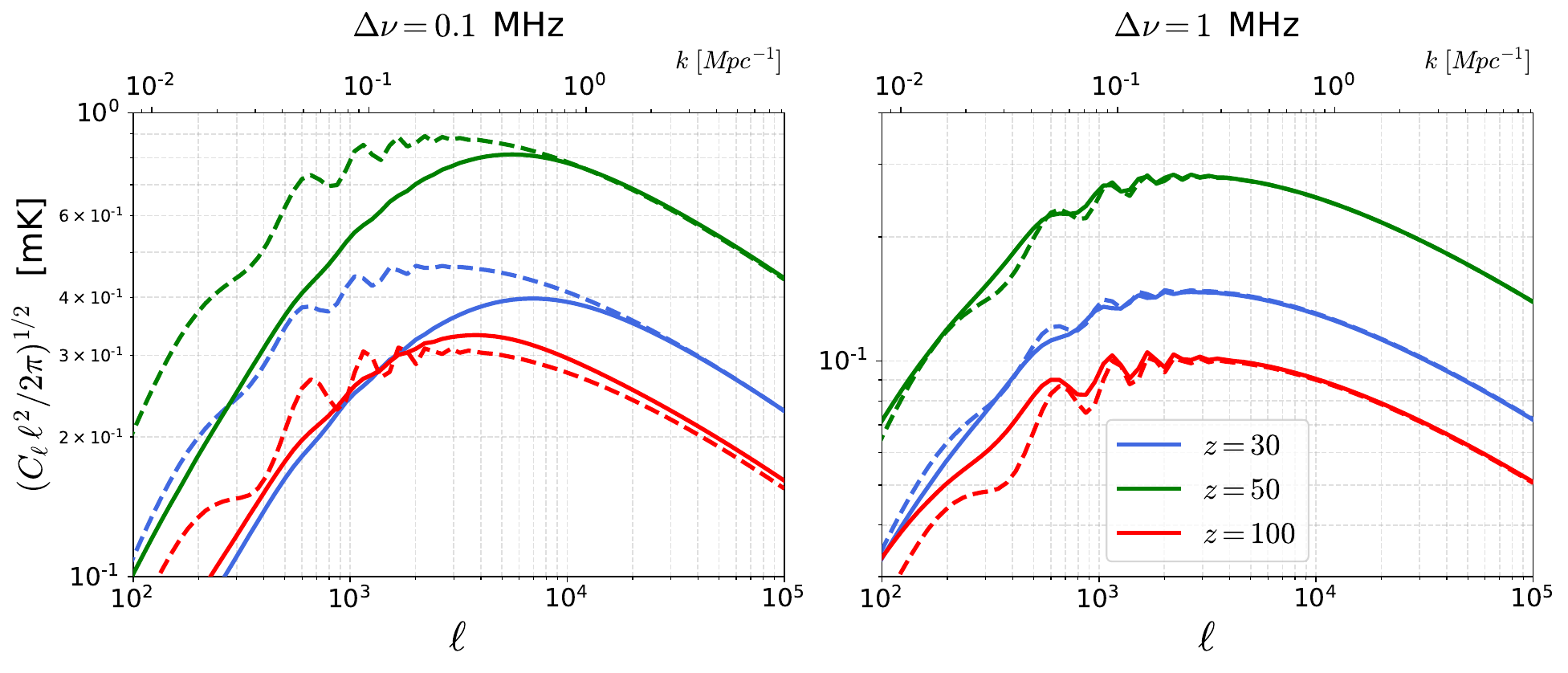} 

    \caption{Power spectrum $C_{\ell}$ of the 21-cm for different values of the redshift and for different choices of the frequency bandwidth of the instrument. Solid lines correspond to the exact flat-sky result including velocity terms. Dashed lines correspond to the Limber approximation where velocity terms are neglected as well. We also show the corresponding co-moving scale associated to each multipole, calculated taking the co-moving distance of emission at $z = 50$.} \label{fig:21cmpower}
\end{figure}

We are now able to compute the resulting 21-cm angular power spectrum at redshift $z$ as
\begin{align} \label{eq:21cmpower}
C^{\rm 21}_{\ell_1 \ell_2}(z) =& \frac{1}{2 \ell_1+1} \sum_{m_1, \, m_2} \, \langle a_{\ell_1 m_1}^{21}(z) \, a_{\ell_2 m_2}^{21 *}(z) \rangle \nonumber \\
=& \delta_{\ell_1 \ell_2} \, 4 \pi \, \int dr' \,  \int dr'' \, W_{r(z)}(r') \, W_{r(z)}(r'') \, \alpha_1(r') \, \alpha_1(r'') \nonumber \\
& \qquad \qquad \qquad \qquad \qquad \times \int \frac{d k}{k} \,\mathcal A_s(k) \, \mathcal{M}_b(k,r') \, \mathcal{M}_b(k,r'') \, j_{\ell}(k r') \, j_{\ell}(k r'') \, .
\end{align}
In this work we are interested in 21-cm anisotropies on very small scales. We can therefore use the Limber approximation to simplify the radial and momentum integrals involving products of spherical Bessel functions. This approximation is valid provided that $\ell_1 \gg r(z)/\delta(r)$, or in other words on scales where the wavelength is much smaller than the redshift bin width (see e.g. \cite{Lewis:2007kz}). According to this approximation, we can assume that the spherical Bessel functions $j_{\ell}(x)$ are small for  $x< \ell$ and peak around $x \sim \ell$. The integral over comoving momenta $k$ will get most of their contribution from modes $k \sim \ell/y$. Therefore, in the Limber approximation we can rewrite the Spherical Bessel functions associated with the 21-cm anisotropies as
\begin{equation} \label{eq:Bessel_Limber}
j_{\ell}(k y) = \sqrt{\frac{\pi}{2 \ell}} \, \delta_{\rm D}(\ell - k y) \, .  
\end{equation}
By substituting Eq. \eqref{eq:Bessel_Limber} into Eq. \eqref{eq:21cmpower}, we can perform the momenta and radial integration using the Dirac deltas. The resulting 21-cm power spectrum simplifies to 
\begin{equation} \label{eq:21cmpower_limber}
C^{\rm 21, Limber}_{\ell_1 \ell_2}(z) = \delta_{\ell_1 \ell_2} \, \frac{2 \pi^2}{\ell_1^3} \, \alpha^2_1(z)   \, \int dy \, y \, \left[W_{r(z)}(y)\right]^2 \,  \, \left[\mathcal{M}_b(\ell_1/y, z)\right]^2 \, \mathcal A_s(\ell_1/y)  \, .  
\end{equation}
An alternative to this approach was proposed in Ref. \cite{Munoz:2015eqa}, where the authors studied 21-cm anisotropies using the flat-sky formalism, including the effect of velocity fluctuations. Here, the momenta are decomposed in terms of the component parallel to the line of sight, \, $k_\parallel = \mathbf{k} \cdot \hat n$,  and the one perpendicular $k_\perp \simeq \ell/r(z)$. The resulting power-spectrum then reads
\begin{equation} \label{eq:21cmpower_flat}
C^{\rm 21, flat-sky}_{\ell_1 \ell_2}(z) = \delta_{\ell_1 \ell_2} \,\frac{1}{r^2(z)} \,\int \frac{d k_{\parallel}}{2 \pi}  \, \left[\tilde W_{r(z)}(k_{\parallel})\right]^2 \, \left[\alpha_1(z) + \overline{T}_{21}(z) \, \frac{k^2_{\parallel}}{k^2} \, \right]^2 \, \left[\mathcal{M}_b(k,z)\right]^2  \,  \mathcal A_s(k)  \, .
\end{equation}
Here $k= \sqrt{k_\parallel^2 + \ell_1^2/r^2(z)}$, $\tilde W_{r(z)}(k_{\parallel})$ is the Fourier transform of the window function introduced in Eq. \eqref{eq:window} and $\overline{T}_{21}(z)$ is the mean 21-cm brightness temperature that enters through the velocity perturbations. In Fig. \ref{fig:21cmpower} we show the expected 21-cm power spectrum at a given redshift slice and for a given angular multipole $\ell_1=|\bsl_1|$. We use both the flat-sky formula in Eq. \eqref{eq:21cmpower_flat} (solid lines) and the Limber approximation in absence of velocity terms, Eq. \eqref{eq:21cmpower_limber} (dashed lines). We employ two choices of $\Delta \nu$, $0.1$ MHz and $1$ MHz. We see that the velocity terms can be neglected at $\ell \gtrsim  10^4$ and $\ell \gtrsim 10^3$, respectively. On scales where the wavelength is much smaller then the redshift bin, $\ell \gtrsim r(z)/\delta(r)$ (which turn out to be the same scales for which the Limber approximation works well), peculiar velocity effects average out. Since $r(z)\sim 10^4 \, \mbox{Mpc}$, $\delta(r) \sim (\Delta \nu/0.1  \mbox{MHz})$ Mpc in the redshift range $z \in [30-100]$, we can neglect peculiar velocity terms at angular scales larger than
\begin{equation} \label{eq:scaling}
\ell_{\rm mono} \gtrsim 10^3/\Delta \nu \, ,
\end{equation}
where $\Delta \nu$ is expressed in MHz. \\

In this work we will be interested in 21-cm anisotropies on co-moving scales $1 \, \mbox{Mpc}^{-1}<k < 10 \, \mbox{Mpc}^{-1}$, within reach of the angular resolution of lunar experiments (see e.g. \cite{Cole:2019zhu}). These correspond roughly to angular scales $\ell \simeq 10^4-10^5 \, \mbox{Mpc}^{-1}$ for which the Limber approximation and neglecting velocity terms lead to robust results, provided the bandwidth of an hypothetical experiment is $\Delta \nu \geq 0.1$ MHz. 

\section{$\langle 21-21- \rm CMB \rangle$ cross-correlation} \label{sec:2121X}

 In this section we compute the contributions to the 21-CMB angular cross-bispectrum\footnote{The two 21-cm fields are assumed to be measured at the same redshift $z$ which we will omit for simplicity of notation.} 
\begin{align} \label{eq:T2121_bis}
\langle a_{\ell_1 m_1}^{21} a_{\ell_2 m_2}^{21} a^X_{\ell_3 m_3}\rangle   \, .
\end{align}
Assuming isotropic fields this can be re-written in terms of the angle-averaged bispectrum $B^{21-21-\rm X}_{\ell_1 \ell_2 \ell_3}$ as
\begin{align} \label{eq:decisot}
\langle a_{\ell_1 m_1}^{21} a_{\ell_2 m_2}^{21} a^X_{\ell_3 m_3}\rangle = \begin{pmatrix}
	\ell_1 & \ell_2 & \ell_3 \\
	m_1 & m_2 & m_3
	\end{pmatrix} \, B^{21-21-\rm X}_{\ell_1 \ell_2 \ell_3} \, .
\end{align}
Note that in general the peculiar velocity terms in the 21-cm field introduce statistical anisotropies, as they depend on the line-of-sight. Therefore, the decomposition \eqref{eq:decisot} can be applied only to the specific scales considered in this work, where we neglect velocity perturbations, resulting in a negligible level of statistical anisotropy. 
Employing Eqs. \eqref{eq:a_CMB_scalar} and \eqref{eq:21_cm_anisotropies} to compute the quantity in Eq. \eqref{eq:T2121_bis}, we find 
\begin{align} \label{eq:cross_full}
\langle a_{\ell_1 m_1}^{21} a_{\ell_2 m_2}^{21} a^X_{\ell_3 m_3}\rangle = & (4 \pi)^3 \, i^{\ell_1 + \ell_2 + \ell_3} \, \int dr' \, \int dr'' \, W_{r(z)}(r') \, W_{r(z)}(r'') \,  \left[\prod_{i=1}^3 \int\frac{d^3 k_i}{(2 \pi)^3} Y_{\ell_i m_i}^*(\hat{k}_i)  \right] \, \nonumber \\
& \quad \times j_{\ell_1}(k_1 r') \, j_{\ell_2}(k_2 r'') \, {\cal T}_{\ell_3(s)}^{X}(k_3) \, \times \langle \dTcm(r'(z),\bk{1}) \, \dTcm(r''(z),\bk{2})  \, \zeta(\bk{3})  \rangle' \nonumber \\
& \quad  \times (2 \pi)^3 \, \delta_{\rm D}^{(3)}(\bk{1} + \bk{2} + \bk{3}) \, . 
\end{align} 
We can expand the Dirac delta as
\begin{align}
\delta_{\rm D}^{(3)}(\bk{1} + \bk{2} + \bk{3}) = 8 \, \int_0^\infty \, y^2 \, dy \, \left[\prod_{n=1}^3 \sum_{L_n M_n} \, (-1)^{L_n/2} \,  j_{L_n}(k_i y) \,Y_{L_n M_n}^*(\hat{k}_i) \right] \, h_{L_1 L_2 L_3}^{0 0 0} \, \begin{pmatrix}
	L_1 & L_2 & L_3 \\
	M_1 & M_2 & M_3
	\end{pmatrix} \, ,
\end{align}
where 
\begin{equation}
h_{\ell_1 \ell_2 \ell_3}^{s_1 s_2 s_3} = \sqrt{\frac{(2 \ell_1 +1)(2 \ell_2 +1)(2 \ell_3 +1)}{4 \pi}} \, \begin{pmatrix}
	\ell_1 & \ell_2 & \ell_3 \\
	s_1 & s_2 & s_3
	\end{pmatrix} \, .
\end{equation}
Eq. \eqref{eq:cross_full} can be written as
\begin{align} \label{eq:cross_full_exp}
\langle a_{\ell_1 m_1}^{21} a_{\ell_2 m_2}^{21} a^X_{\ell_3 m_3}\rangle = & \frac{8}{\pi^3} \, i^{\ell_1 + \ell_2 + \ell_3} \, \int dr' \, \int dr'' \, W_{r(z)}(r') \, W_{r(z)}(r'') \, \int_0^\infty \, y^2 \, dy \, \left[\prod_{i=1}^3 \int d^3 k_i \, Y_{\ell_i m_i}^*(\hat{k}_i)  \right] \, \nonumber \\
& \quad \times \left[\prod_{n=1}^3 \sum_{L_n M_n} \, (-1)^{L_n/2} \,  j_{L_n}(k_i y) \,Y_{L_n M_n}^*(\hat{k}_i) \right] \, h_{L_1 L_2 L_3}^{0 0 0} \, \begin{pmatrix}
	L_1 & L_2 & L_3 \\
	M_1 & M_2 & M_3
	\end{pmatrix}  \nonumber \\
& \quad \times j_{\ell_1}(k_1 r') \, j_{\ell_2}(k_2 r'') \, {\cal T}_{\ell_3(s)}^{X}(k_3) \, \times \langle \dTcm(r'(z),\bk{1}) \, \dTcm(r''(z),\bk{2})  \, \zeta(\bk{3})  \rangle'  \, . 
\end{align} 
Next, we must specify a form for the cross-correlated $\langle \rm 21-21-\zeta \rangle$ bispectrum in Fourier space.

\subsection{Primordial contribution} \label{sub:prim}

Taking the local ansatz of Eq. \eqref{eq:BLocal}, the primordial contribution to the $\langle \rm 21-21-\zeta \rangle$ cross-correlated bispectrum in Fourier space reads
\begin{align} \label{eq:bisp_prim_scal}
 \langle \dTcm(r'(z),\bk{1}) \, \dTcm(r''(z),\bk{2})  \, \zeta(\bk{3})  \rangle'_{\rm prim}  = & \langle \dTcm^{(1)}(r'(z),\bk{1}) \, \dTcm^{(1)}(r''(z),\bk{2})  \, \zeta(\bk{3})  \rangle' \nonumber \\
 = & \alpha_1(r') \, \alpha_1(r'') \,  \mathcal{M}_b(k_{1}, r') \,  \mathcal{M}_b(k_{2}, r'') \, \times B^{\text{loc}}_{\zeta\zeta\zeta}(k_{1},k_{2},k_{3}) \, .
\end{align}
Inserting Eq. \eqref{eq:bisp_prim_scal} into Eq. \eqref{eq:cross_full_exp} we get the following final expression for the primordial contribution to the angular-averaged bispectrum (see App. \ref{app:computations_prim} for details)
\begin{align} \label{eq:cross_full_prim_tot}
B^{21-21-\rm X, \, prim}_{\ell_1 \ell_2 \ell_3}(z) = & \, \fnl{loc} \, \frac{192 \pi}{5} \, (-1)^{\ell_1 + \ell_2 + \ell_3} \, \sqrt{\frac{(2 \ell_1+1) (2 \ell_2+1) (2 \ell_3+1)}{(4 \pi)}} \, \begin{pmatrix}
	\ell_1 & \ell_2 & \ell_3 \\
	0 & 0 & 0
	\end{pmatrix}   \nonumber \\
 & \qquad \qquad \times  \Big(\mathcal I^{\rm prim, 1}_{\ell_1 \ell_2 \ell_3} + \mathcal I^{\rm prim, 2}_{\ell_1 \ell_2 \ell_3} \Big) \, ,
\end{align}
where $\mathcal I^{\rm prim, 1}_{\ell_1 \ell_2 \ell_3}$ and $\mathcal I^{\rm prim, 2}_{\ell_1 \ell_2 \ell_3}$ are given in Eqs. \eqref{eq:I1prim} and \eqref{eq:I2prim}, respectively.

\subsection{Secondary contribution} \label{sub:sec}

In addition to the primordial contribution, there are higher order, secondary contributions coming from the second order expression of the 21-cm fluctuation in Eq. \eqref{eq:dT21expansion}\footnote{In principle, there can also be a non-zero contribution coming from second-order effects in the CMB anisotropies, which however can be neglected (see e.g. \cite{Hanson:2009kg} for this kind of studies in the CMB bispectrum context).},
\begin{align}\label{eq:secbispectrum}
	\langle \dTcm &(\bk{1})\dTcm(\ktwo)\zeta(\kthree)\rangle_{\mathrm{sec.}}' = \langle \dTcm^{(2)} (\bk{1})\dTcm^{(1)}(\ktwo) \zeta(\kthree)\rangle' +1\;\mathrm{perm}\nonumber\\
	&= 2  \, \alpha_1 \, \alpha_2 \,{F}_2^{(s)}(\bk{2},\bk{3}) \, P_\delta(\bk{2}) \, P_\delta(\bk{3}) \, \left[\mathcal{M}_b(\bk{3})\right]^{-1} \nonumber\\
	&\qquad +2 \, \alpha_1 \, \alpha_3 \, P_\delta(\bk{2}) \, P_\delta(\bk{3}) \, \left[\mathcal{M}_b(\bk{3})\right]^{-1} +1\;\mathrm{perm \, ,} 
\end{align}
where we have used Eq. \eqref{eq:dT21expansion}. 
We obtain to the final result (see App. \ref{app:computations_sec} for the full derivation) 
\begin{align} \label{eq:cross_full_secfinal}
& B^{21-21-\rm X, \, sec}_{\ell_1 \ell_2 \ell_3}(z) = 64 \pi \,  \sqrt{\frac{(2 \ell_1+1) (2 \ell_2+1) (2 \ell_3+1)}{4 \pi}} \,    \nonumber \\
 & \times  \Bigg[(-1)^{\ell_1 + \ell_2 + \ell_3} \, \begin{pmatrix}
	\ell_1 & \ell_2 & \ell_3 \\
	0 & 0 & 0
	\end{pmatrix} \,  \mathcal I^{\rm sec, 1}_{\ell_1 \ell_2 \ell_3}   \nonumber \\
& \qquad \qquad - c_2  \, i^{\ell_2 + \ell_3}  \, \sum_{L_2 L_3} \, i^{L_2 + L_3} \, (2 L_2 +1) \, (2 L_3 +1) \, \, \begin{pmatrix}
	\ell_1 & L_2 & L_3 \\
	0 & 0 & 0
	\end{pmatrix} \, \begin{pmatrix}
	\ell_2 & L_2 & 1 \\
	0 & 0 & 0
	\end{pmatrix}  \, \begin{pmatrix}
	\ell_3 & L_3 & 1 \\
	0 & 0 & 0
	\end{pmatrix}  \, \left\{\begin{matrix}
	\ell_1 & \ell_2 & \ell_3 \\
	1 & L_3 & L_2	\end{matrix}\right\} \,   \nonumber \\
& \qquad \qquad \qquad \qquad \times \left(\mathcal I^{\rm sec, 2}_{\ell_1 \ell_2 \ell_3, \, L_2 L_3} + \mathcal I^{\rm sec, 3}_{\ell_1 \ell_2 \ell_3, \, L_2 L_3}\right) \nonumber \\
& \qquad \qquad + \frac{2}{3}  c_3 \, i^{\ell_2 + \ell_3} \sum_{L_2 L_3} \, i^{L_2 + L_3} \, (2 L_2 +1) \, (2 L_3 +1) \, \, \begin{pmatrix}
	\ell_1 & L_2 & L_3 \\
	0 & 0 & 0
	\end{pmatrix} \, \begin{pmatrix}
	\ell_2 & L_2 & 2 \\
	0 & 0 & 0
	\end{pmatrix}  \, \begin{pmatrix}
	\ell_3 & L_3 & 2 \\
	0 & 0 & 0
	\end{pmatrix}  \, \left\{\begin{matrix}
	\ell_1 & \ell_2 & \ell_3 \\
	2 & L_3 & L_2	\end{matrix}\right\} \nonumber  \\
&  \qquad \qquad \qquad \qquad  \times \left(\mathcal I^{\rm sec, 4}_{\ell_1 \ell_2 \ell_3, \, L_2 L_3}\right) \, \Bigg] \nonumber \\
& \qquad \qquad \qquad \qquad \qquad \qquad \qquad \qquad \qquad \qquad  + \ell_1 \leftrightarrow \ell_2 \, ,
\end{align}
where  $\mathcal I^{\rm sec, 1}_{\ell_1 \ell_2 \ell_3}-\mathcal I^{\rm sec, 4}_{\ell_1 \ell_2 \ell_3, \, L_2 L_3}$ are given in Eqs. \eqref{eq:I1limberm}-\eqref{eq:I4limberm}.
\begin{figure}[h!]
        \centering
        \includegraphics[width=\linewidth]{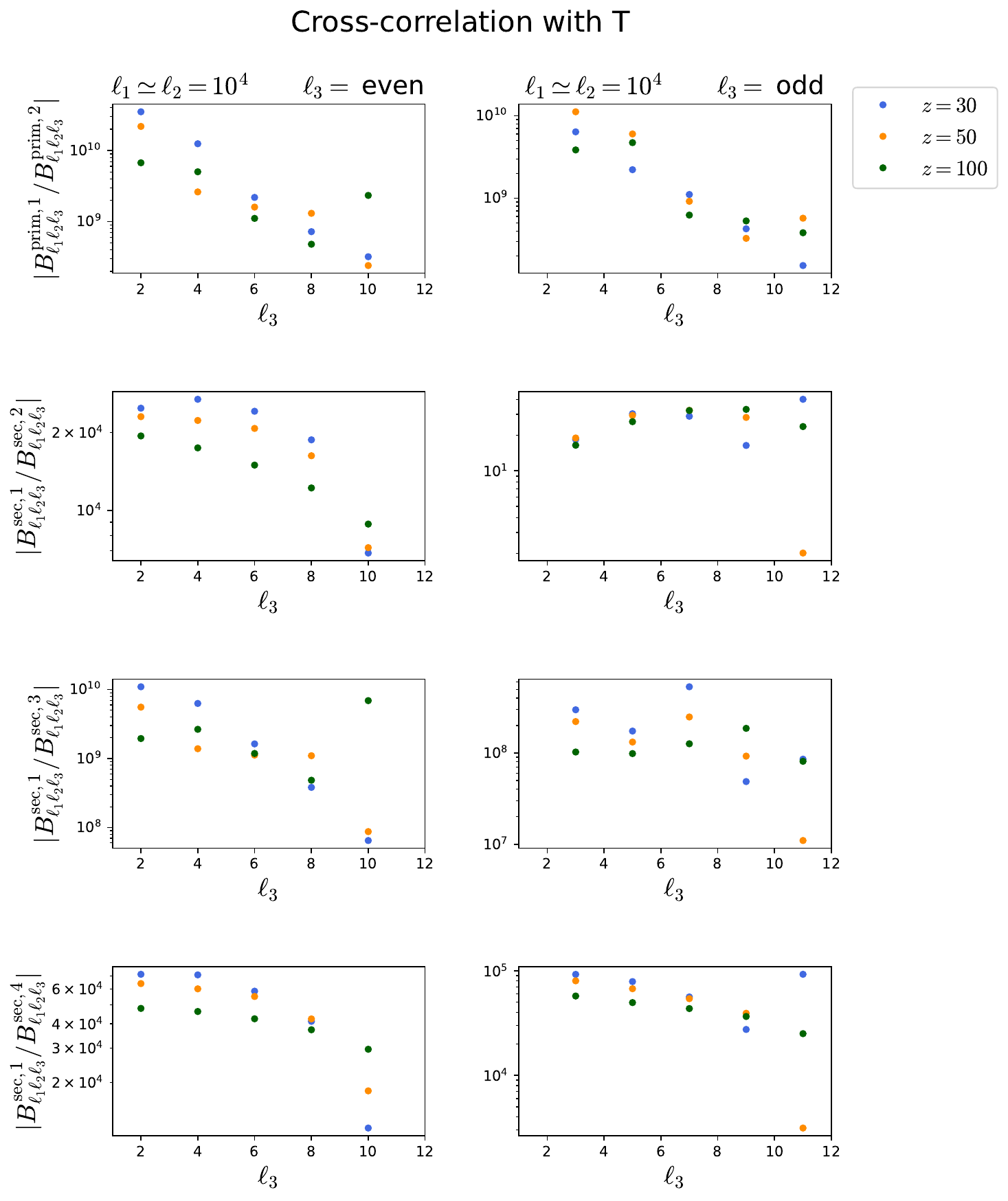} 
    \caption{Ratios between the leading terms $\mathcal I^{\rm prim, 1}_{\ell_1 \ell_2 \ell_3}$, $\mathcal I^{\rm sec, 1}_{\ell_1 \ell_2 \ell_3}$ and the other terms contributing to the primordial and secondary $\langle 21 - 21 - T \rangle$ bispectrum.} \label{fig:match_prim_T}
\end{figure}
\begin{figure}[h!]
        \centering
        \includegraphics[width=\linewidth]{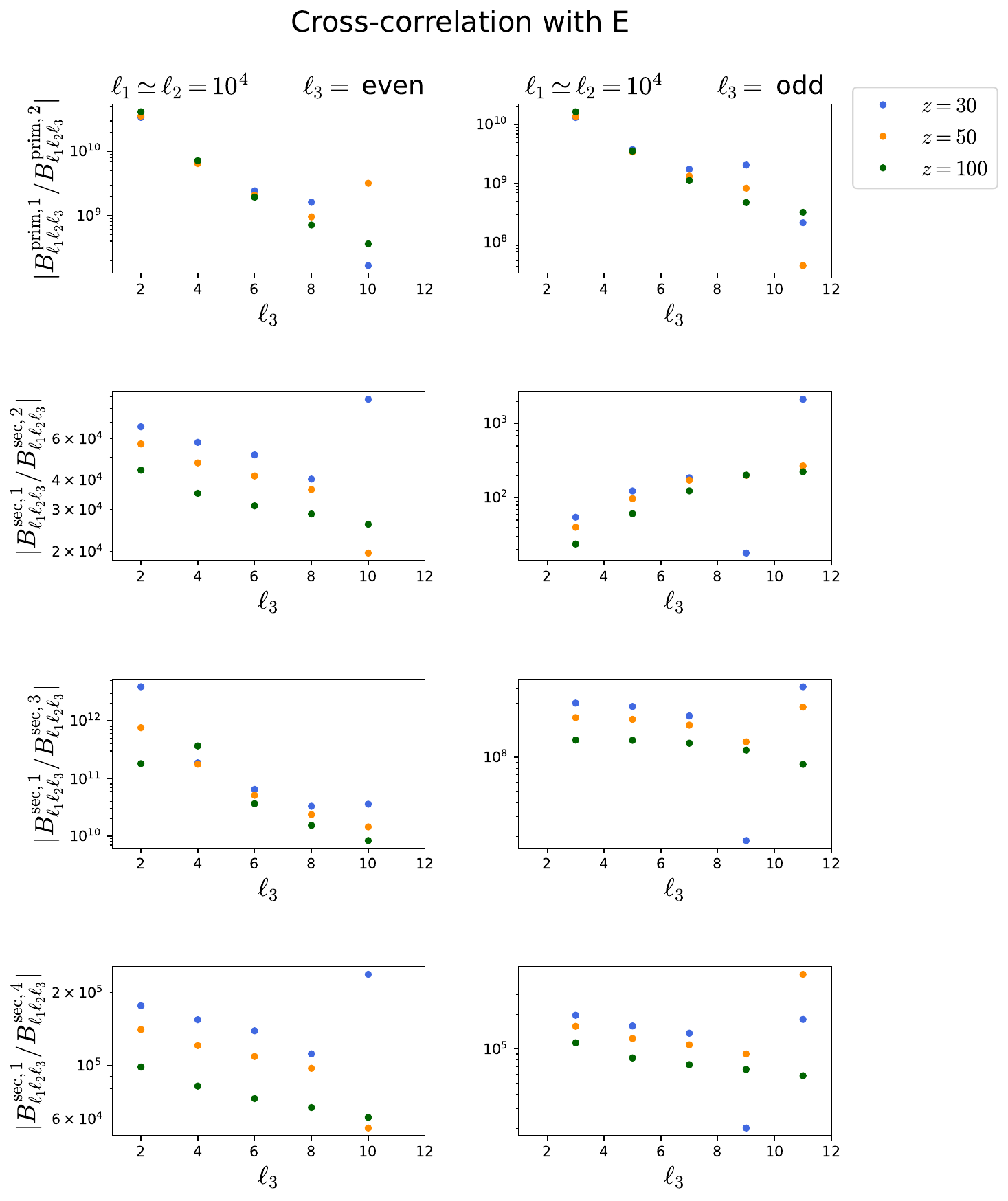} 
    \caption{Same as Fig. \ref{fig:match_prim_T} for the signal coming from cross-correlations with CMB $E$ modes.} \label{fig:match_prim_E}
\end{figure}

\subsection{Primordial-secondary comparison}

By inspecting the $\ell_1, \ell_2$ ($> 10^4$) and $y$ ($\simeq y(z)$) scalings of the equations for the primordial and secondary contributions we expect that the terms $\mathcal I^{\rm prim, 1}_{\ell_1 \ell_2 \ell_3}$,  Eq. \eqref{eq:I1prim_Limber} and $\mathcal I^{\rm sec, 1}_{\ell_1 \ell_2 \ell_3}$, Eq. \eqref{eq:I1limberm} dominate the primordial and secondary contributions respectively. In Figs. \ref{fig:match_prim_T} and \ref{fig:match_prim_E} we show the ratios between the $\mathcal I^{\rm prim, 1}$ and $\mathcal I^{\rm sec, 1}$ contributions with respect to the other contributions in the squeezed configurations (these contain most of the Fisher information, as we will show below) for the CMB $T$- and $E$-mode. The only other term which appears to be somewhat relevant is $\mathcal I^{\rm sec, 2}$ for $\ell_3 = \rm odd$. Therefore, to evaluate quantities within sub-percent error it is sufficient to consider $\mathcal I^{\rm prim, 1}$, $\mathcal I^{\rm sec, 1}$ and $\mathcal I^{\rm sec, 2}$ and neglect all the other terms.

\begin{figure}[h!]
        \centering
        \includegraphics[width=\linewidth]{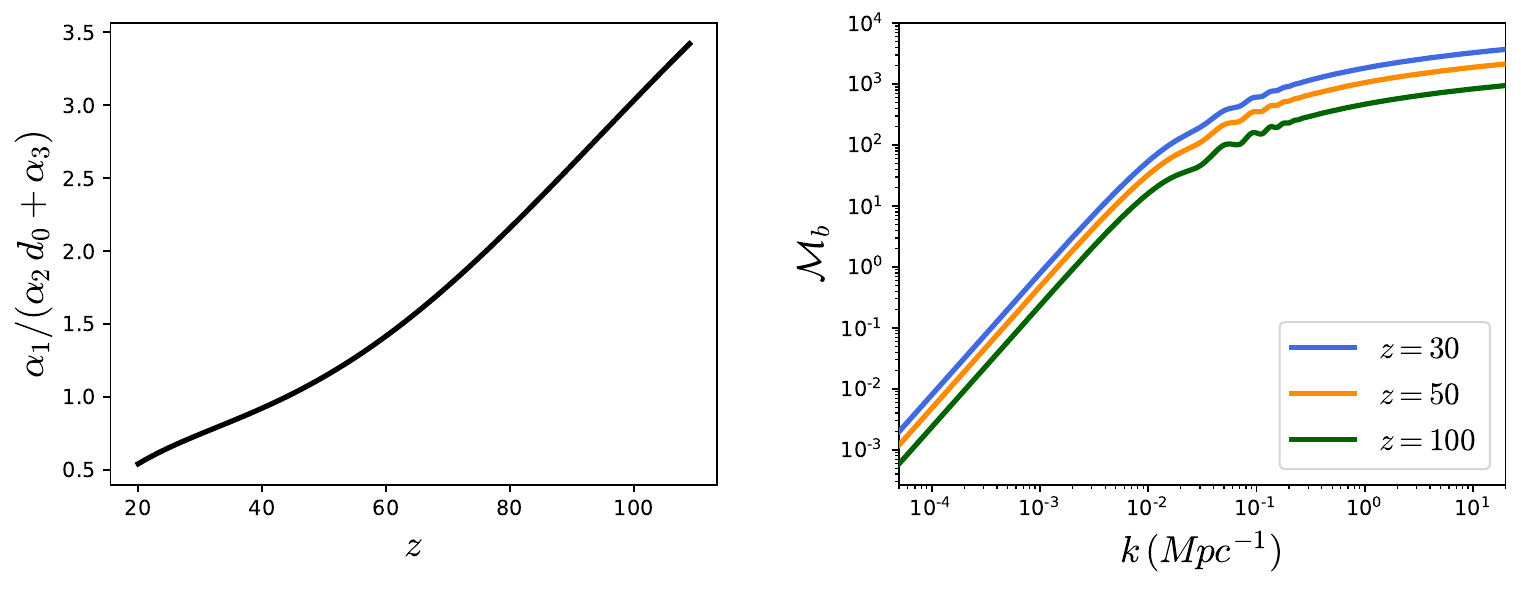} 
    \caption{Left: redshift dependent factor as appearing in Eq. \eqref{eq:estiationprimoversec}. Right: Baryon transfer function for different redshifts.} \label{fig:est}
\end{figure}

Given the fact that $\mathcal I^{\rm prim, 1}$ and $\mathcal I^{\rm sec, 1}$ are the dominant contributions we can roughly estimate the ratio between the primordial and secondary contributions at a given redshift slice $z$ as
\begin{equation} \label{eq:estiationprimoversec}
\frac{B^{21-21-\rm X, \, prim}_{\ell_1 \ell_2 \ell_3}(z)}{B^{21-21-\rm X, \, sec}_{\ell_1 \ell_2 \ell_3}(z)} \approx 0.6 \, \fnl{loc} \, \left[\frac{\alpha_1(z)}{\alpha_2(z) \, d_0 +  \alpha_3(z)}\right] \, \mathcal M^{-1}_b\left(\tilde k_3\right) \, ,
\end{equation}
where $\tilde k_3$ refers to the characteristic scale which mostly contributes to the CMB $\ell_3$-pole anisotropies of a given mode $X$. As the redshift dependent factor in Eq. \eqref{eq:estiationprimoversec} is of order $1$ and the CMB largest scales correspond to $\tilde k_3 \simeq 10^{-4} \, \mbox{Mpc}^{-1}$ for which $\mathcal M_b\left(\tilde k_3\right) \ll 1$ (see Fig.~\ref{fig:est}), it follows that in our specific case the secondary contributions represent a contamination to the primordial signal much less problematic than that found in studies of the 21-cm auto-bispectrum, where secondaries systematically dominate the primordial signals by $2$ to $3$ orders of magnitude for $\fnl{loc} = 1$ (see e.g. Ref. \cite{Munoz:2015eqa}). 

\begin{figure}[h!]
        \centering
        \includegraphics[width=0.97\linewidth]{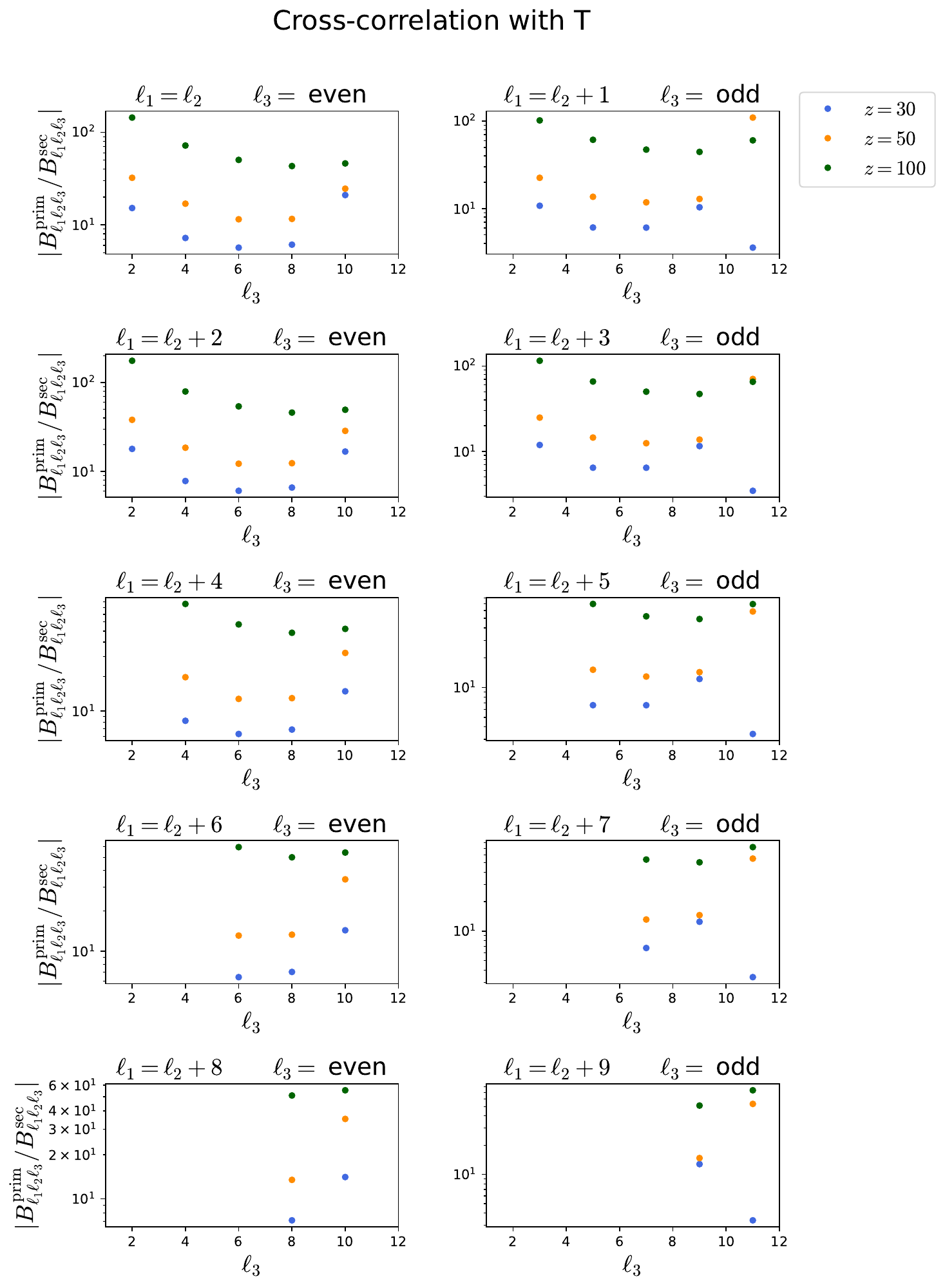} 
    \caption{Ratios between primordial and secondary contributions for different multipole configurations and redshift slices. $f_{\rm NL}^{\rm loc} = 1$ is assumed.} \label{fig:prim_sec}
\end{figure}

\begin{figure}[h!]
        \centering
        \includegraphics[width=0.97\linewidth]{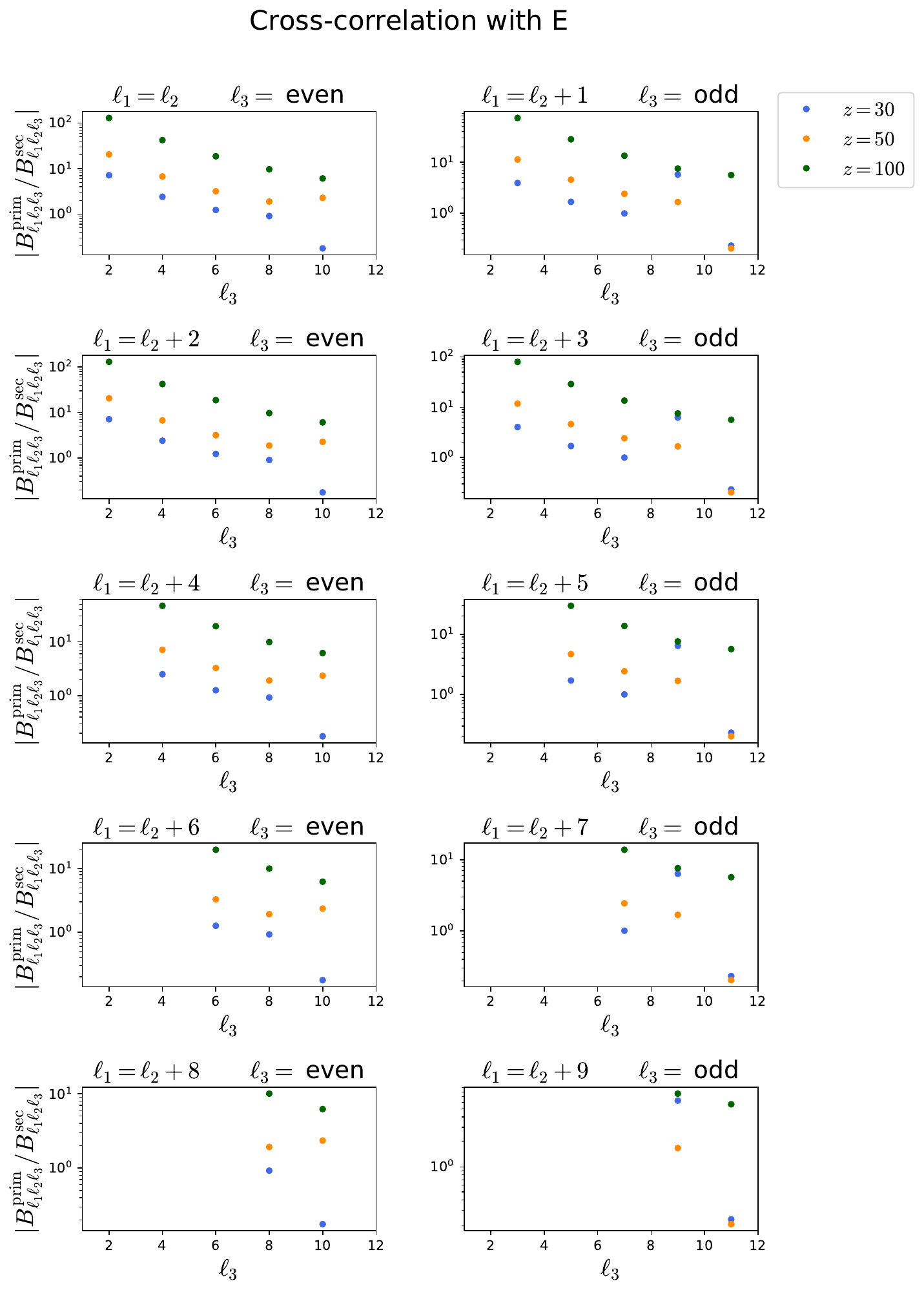} 
    \caption{Same as Fig. \ref{fig:prim_sec} for the cross-correlation with the CMB $E$-polarization mode.} \label{fig:prim_secpol}
\end{figure}

In Figs. \ref{fig:prim_sec} and \ref{fig:prim_secpol} we show the exact ratios between the primordial ($f_{\rm NL}^{\rm loc} = 1$) and secondary contributions for different multipole configurations and redshift slices. We focus on the cross-correlations of the 21-cm fluctuations with either the CMB $T$- (Fig. \ref{fig:prim_sec}) and $E$-mode (Fig. \ref{fig:prim_secpol}) sourced by scalar perturbations. We conclude that up to $\ell_3 \simeq 10$ and for all the possible configurations allowed by the triangle of momenta, the secondary contributions are subdominant to the primordial signal, with up to $2$ orders of magnitude difference at the largest CMB scale $\ell_3= 2$. 
A similar result is found also for cross-correlations with the CMB polarization field $E$. Here however the primordial signal tends to be less dominant with respect to the secondary, with some multipole configurations even dominated by the secondary signal. We can understand this by realizing that for a fixed multipole $\ell_3$ $E$-mode CMB anisotropies receive contributions from scales smaller than the $T$-mode counterpart, resulting in larger values of $\mathcal M_b\left(\tilde k_3\right)$.

\FloatBarrier
\section{Fisher analysis} \label{sec:fisher}

\subsection{Single redshift slice} \label{sub:singlered}

In this section we investigate the constraining power of cross-correlations on $\fnl{loc}$ by means of a Fisher analysis. We start by considering the cross-correlation of two 21-cm anisotropies and one $T$-mode CMB anisotropy.  Later on we will implement the information coming from the polarization of the CMB. Furthermore, we neglect secondary contributions which we will consider in the following. Assuming that we have made an observation of the $\langle \rm 21-21-T\rangle$ cross-correlation, the minimum-variance null-hypothesis (MVNH) estimator for $f_{\rm NL}$ at a single redshift slice $z$ is given by (see e.g. \cite{Babich:2004yc,Smith:2011rm,Smith:2012ta})
\begin{equation} \label{eq:estimatorfNL}
\hat f_{\rm NL}|_z = \frac{1}{F(z)} \sum_{\ell_3 \leq \ell_2 \leq \ell_1} \, \frac{B^{21-21-\rm T, \, obs}_{\ell_1 \ell_2 \ell_3}(z) \,\,\, \tilde B^{21-21-\rm T, \, prim}_{\ell_1 \ell_2 \ell_3}(z)}{C^{21}_{\ell_1}(z) \, C^{21}_{\ell_2}(z) \, C^{\rm T}_{\ell_3}} \, ,
\end{equation}
where $B^{21-21-\rm T, \, obs}_{\ell_1 \ell_2 \ell_3}(z)$ is the following unbiased estimator for the angle-averaged $\langle \rm 21-21-T\rangle$ bispectrum
\begin{equation} 
B^{21-21-\rm T, obs}_{\ell_1 \ell_2 \ell_3}(z) = \sum_{m_1 m_2 m_3} \, \begin{pmatrix}
	\ell_1 & \ell_2 & \ell_3 \\
	m_1 & m_2 & m_3
	\end{pmatrix} \, a_{\ell_1 m_1}^{21, \rm obs}(z) \, a_{\ell_2 m_2}^{21, \rm obs}(z) \, a^{T, \rm obs}_{\ell_3 m_3} \, ,
\end{equation}
$\tilde B^{21-21-\rm T, \, prim}_{\ell_1 \ell_2 \ell_3}$ is defined such that 
\begin{equation}
B^{21-21-\rm T, \, prim}_{\ell_1 \ell_2 \ell_3}  = \fnl{loc} \, \tilde B^{21-21-\rm T, \, prim}_{\ell_1 \ell_2 \ell_3}  \, . 
\end{equation}
The normalization $F(z)$ is the Fisher information for $f_{\rm NL}$ which is given by
\begin{equation} \label{eq:Fisher}
F(z) = f_{\rm sky} \, \sum_{\ell_3 \leq \ell_2 \leq \ell_1} \, \frac{\left(\tilde B^{21-21-\rm T, \, prim}_{\ell_1 \ell_2 \ell_3}(z)\right)^2}{\Delta_{\ell_1 \ell_2} \, C^{21}_{\ell_1}(z) \, C^{21}_{\ell_2}(z) \, C^{\rm T}_{\ell_3}} \, ,
\end{equation}
where $\Delta_{\ell_1 \ell_2} = 2$ if $\ell_1 = \ell_2$, $\Delta_{\ell_1 \ell_2} = 1$ if $\ell_1 \neq \ell_2$,  and $f_{\rm sky}$ is the fraction of the sky covered by the joint-experiments. The sums over the multipoles run over the triangular configurations 
\begin{equation}
\ell_3 \geq |\ell_1 - \ell_2| \,, \qquad \qquad \ell_3 \leq \ell_1 + \ell_2 \, ,
\end{equation}
and are non-zero only when $\ell_1 + \ell_2 + \ell_3 = \mbox{even}$, since the signal is parity-invariant. 

In principle, the 21-cm and $T$-mode CMB power spectra in the denominator of Eq. \eqref{eq:Fisher} should contain cosmic variance and additional sources of noise, including instrumental noise and foregrounds. Here we assume a cosmic variance limited detection of CMB and 21-cm anisotropies. Our analysis further assumes that the covariance of the primordial signal can be approximated by the diagonal Gaussian contribution, neglecting any non-Gaussian covariance. We will comment on these assumptions in the final discussion. 

\begin{figure}[h!]
        \centering
        
        \includegraphics[width=0.49\textwidth]{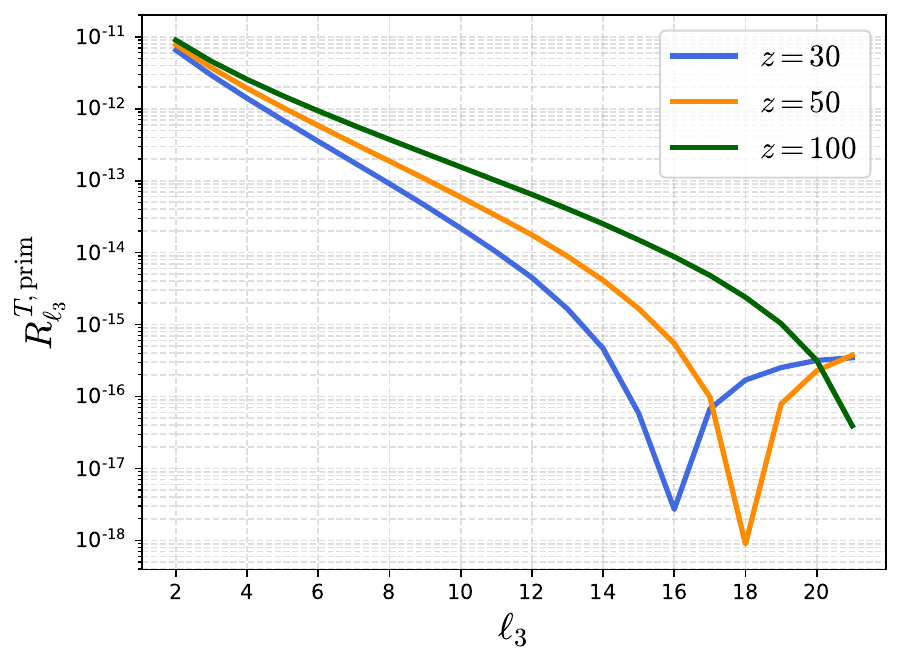} 
\caption{$\ell_3$ dependence of each term in the Fisher matrix summation, Eq. \eqref{eq:RTpriml3}.} \label{fig:RTpriml3}
\end{figure}

The cosmic-variance limited CMB and 21-cm power spectra are given by Eq. \eqref{eq:Xspectrum} and \eqref{eq:21cmpower}, respectively. It is interesting to evaluate the ratio between the squared $k_3$-integral appearing in Eq. \eqref{eq:I2prim} at a given redshift $z$ versus the CMB temperature power spectrum
\begin{equation}  \label{eq:RTpriml3}
R^{T, \, \rm prim}_{\ell_3}(z) = \frac{\left[\int \frac{d k_3}{k_3} \, j_{\ell_3}(k_3 r(z)) \, {\cal T}_{\ell_3(s)}^{T}(k_3) \,\mathcal A_s(k_3) \right]^2}{C^{\rm T}_{\ell_3}}  \, . 
\end{equation}
This factor gives us information about the $\ell_3$ dependence of each term in the Fisher matrix summation. From Fig. \ref{fig:RTpriml3} it is evident that for fixed values of the pair $\ell_1, \ell_2$, each term in the $\ell_3$ summation decreases relatively fast, in such a way that we expect that only the triangular configurations where $\ell_3$ is small ($\lesssim 10$) and $\ell_1 \simeq \ell_2$ contribute significantly to the Fisher matrix. This is explained physically as on large scales CMB $T$-mode anisotropies are mostly sourced during recombination ($z \simeq 1100$), much before the Dark Ages ($z \simeq 30-100$). The different emission time causes the projection of a given primordial signal into different CMB and 21-cm multipole scales and vice-versa, according to the scaling 
\begin{equation} \label{eq:multipoletomomentum}
\ell_3 \propto k_3 \, r(z) \, .
\end{equation}
Given the fact that the CMB is emitted before the 21-cm signal, the same multipole scale $\ell_3$ is affected by physical scales that for the 21-cm are slightly smaller than the CMB. Therefore, on large (CMB) scales the resulting cross-correlation will receive a slight damping compared to the case where if the two signals are emitted at the same time. By increasing $\ell_3$ the CMB transfer function ${\cal T}_{\ell_3(s)}^{T}(k_3)$ becomes more and more narrow in the momentum space, increasing this damping effect and decreasing the power of the cross-correlation with respect to the CMB power spectrum. This results in the relatively fast drop as a function of $\ell_3$ as seen in Fig. \ref{fig:RTpriml3}. This effect is expected to be less severe at redshifts closer to recombination, as we confirm in the same figure.

\begin{figure}
        \centering
        T only\par\medskip
        \includegraphics[width=.49\textwidth]{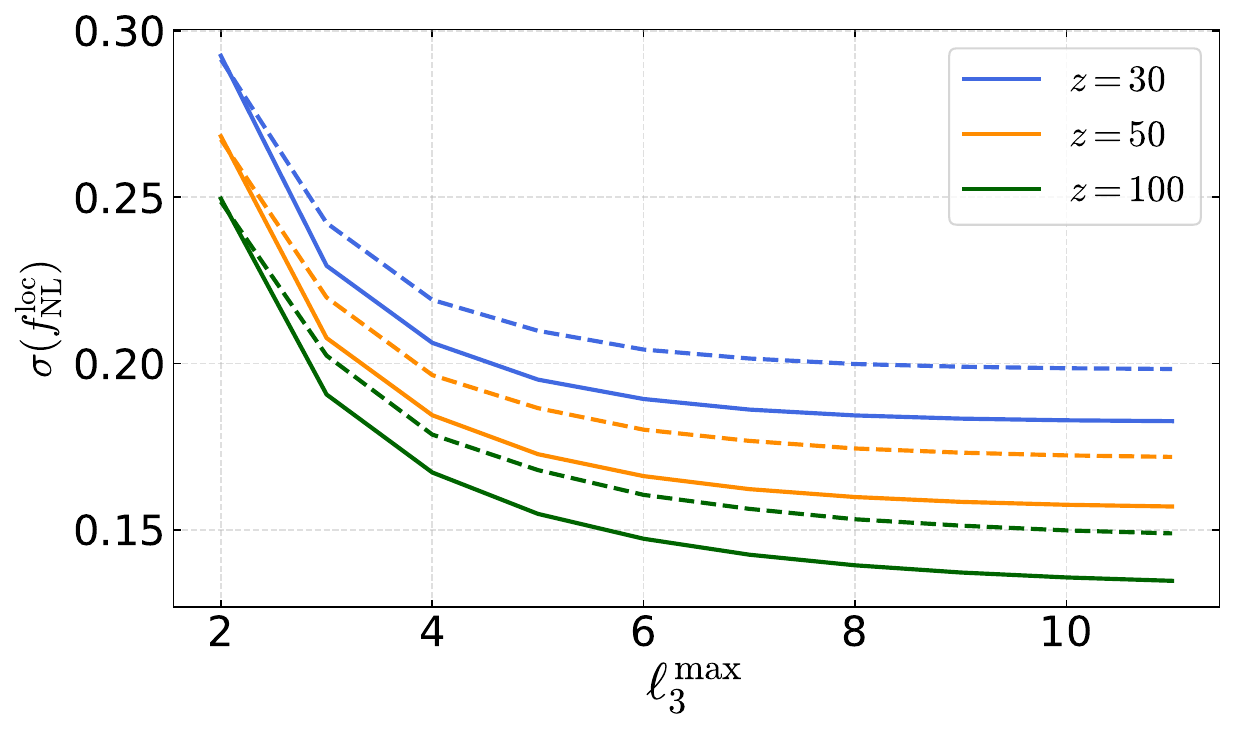} 
        \includegraphics[width=.49\textwidth]{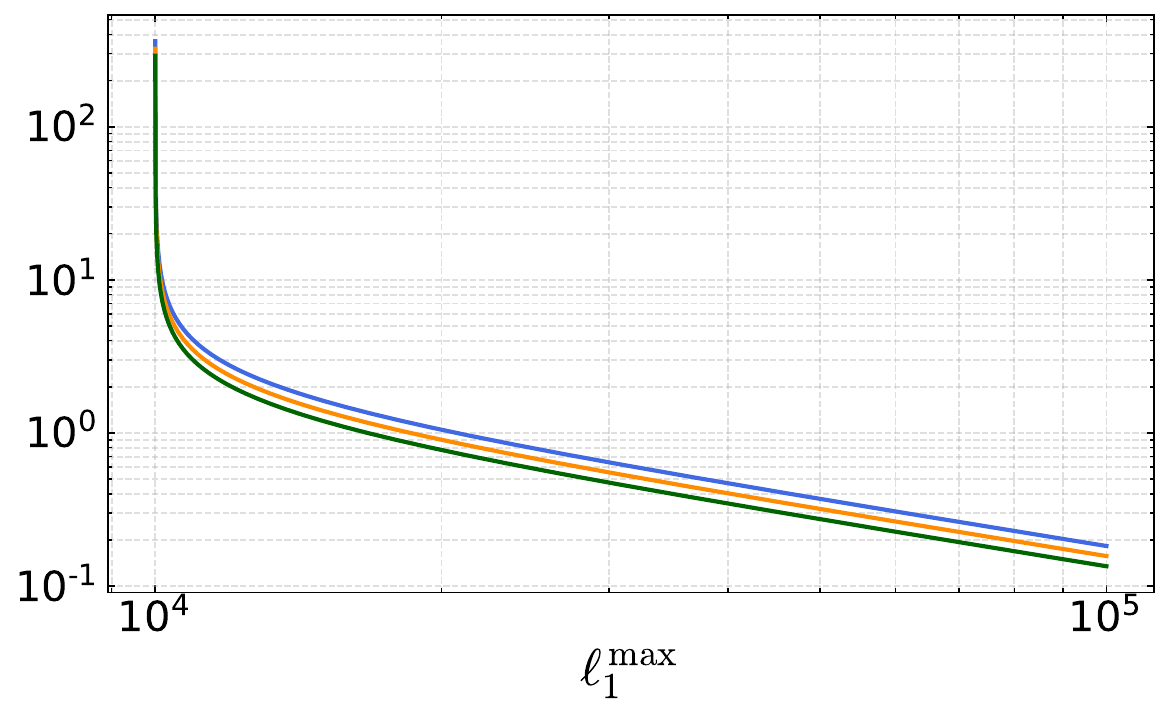}  
    \caption{Expected 1$\sigma$ uncertainty on $\fnl{loc}$ per single redshift slice, using only the CMB $T$-modes. Left panel: as a function of the maximum CMB multipole $\ell^{\rm max}_3$ and summing $\ell_1$ and $\ell_2$ over the $[10^4-10^5]$ domain. Dashed lines correspond to the estimation using the approximated scaling formula \eqref{eq:SNR_approx}. Right panel: as a function of $\ell^{\rm max}_1$ and summing $\ell_3$ over the range $[2-11]$.} \label{fig:forecastfNLT}
\end{figure}
By combining Eqs. \eqref{eq:Xspectrum}, \eqref{eq:21cmpower} and \eqref{eq:cross_full_prim_tot}, and taking $\ell_1 \simeq \ell_2$, we get the following approximated scaling formula for the Fisher matrix
\begin{align} \label{eq:Fisher_approx}
F(z) \approx  \, 180 \, f_{\rm sky}  \, \left[\sum_{\ell_1} \, \ell_1\right] \, \left[\sum_{\ell_3} R_{\ell_3}(z) \, N_{\ell_3} \right] = \, 9 \times 10^{11} \, f_{\rm sky} \, \left( \frac{\ell_{1, \rm max}}{10^5}\right)^2  \, \left[\sum_{\ell_3} R_{\ell_3}(z) \, N_{\ell_3} \right]  \, ,
\end{align}
where the $\ell_3$-dependent quantity $N_{\ell_3}$  is the number of $\ell_2$ multipoles that are smaller or equal than $\ell_1$ and greater than $|\ell_1-\ell_3|$ for a given $\ell_1$. 
The resulting 1$\sigma$ uncertainty in measuring $\fnl{loc}$ is 
\begin{equation} \label{eq:SNR_approx}
\sigma(f_{\rm NL}^{\rm loc})_T(z) = \left[F(z)\right]^{-1/2} \approx 1.05 \times 10^{-6}  \,  f^{-1/2}_{\rm sky} \, \, \left( \frac{10^5}{\ell_{1}^{\rm max}}\right) \, \left[\sum_{\ell_3} R_{\ell_3}(z) \, N_{\ell_3} \right]^{-1/2} \, . 
\end{equation}
In Fig. \eqref{fig:forecastfNLT} we show the expected uncertainty in the measurement of $\fnl{loc}$ for different redshift slices assuming a full-sky experiment ($f_{\rm sky} = 1$), using both the exact expression for the Fisher matrix \eqref{eq:Fisher} and the approximated scaling derived in Eq. \eqref{eq:SNR_approx}. We fix $\ell^{\rm min}_1 = \ell^{\rm min}_2 =  10^4, \, \ell^{\rm min}_3 =  2$ and $\ell^{\rm max}_1 = \ell^{\rm max}_2 =  10^5, \,  \ell^{\rm max}_3 =  11$. We see that the approximated formula \eqref{eq:SNR_approx} works relatively well when $\ell^{\rm max}_3 =  2$, while it becomes less accurate at higher $\ell^{\rm max}_3$, though still providing the correct order of magnitude. As expected, we see a saturation of the Fisher information at $\ell^{\rm max}_3 \approx 10$. Any information from the CMB on smaller scales does not lead to a significant improvement in the measurement of $\fnl{loc}$. Furthermore, from the bottom panel of Fig. \ref{fig:forecastfNLT} we can derive the following scaling of the uncertainty as a function of the smallest 21-cm scale 
\begin{equation} \label{eq:fNLT}
\sigma(f_{\rm NL}^{\rm loc})_T\simeq 0.2  \, f^{-1/2}_{\rm sky} \, \left( \frac{10^5}{\ell_{1}^{\rm max}}\right) \, .
\end{equation}\\

\begin{figure}    
        \centering
        T+E \par\medskip
        \includegraphics[width=0.49\textwidth]{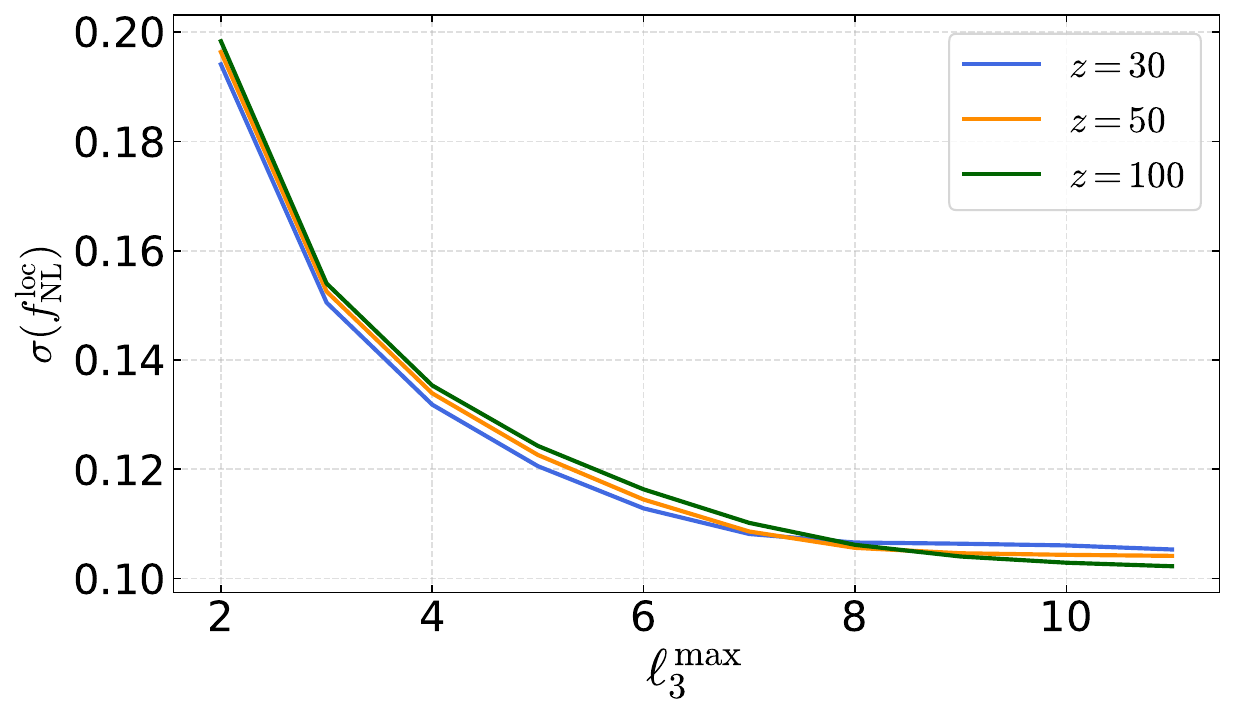} 
        \includegraphics[width=0.49\textwidth]{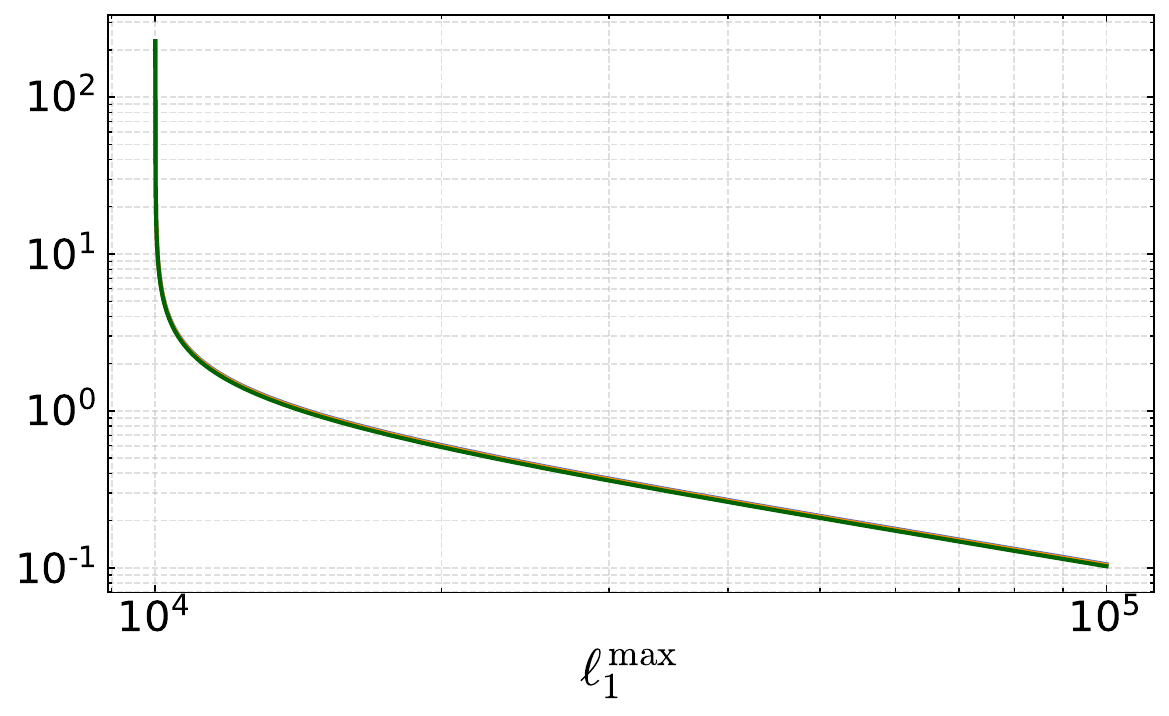}  
    \caption{Same as Fig. \ref{fig:forecastfNLT} but with the addition of the CMB polarization field.} \label{fig:forecastfNLTpE}
\end{figure}

\begin{figure}[h!]
        \centering
        \includegraphics[width=0.49\textwidth]{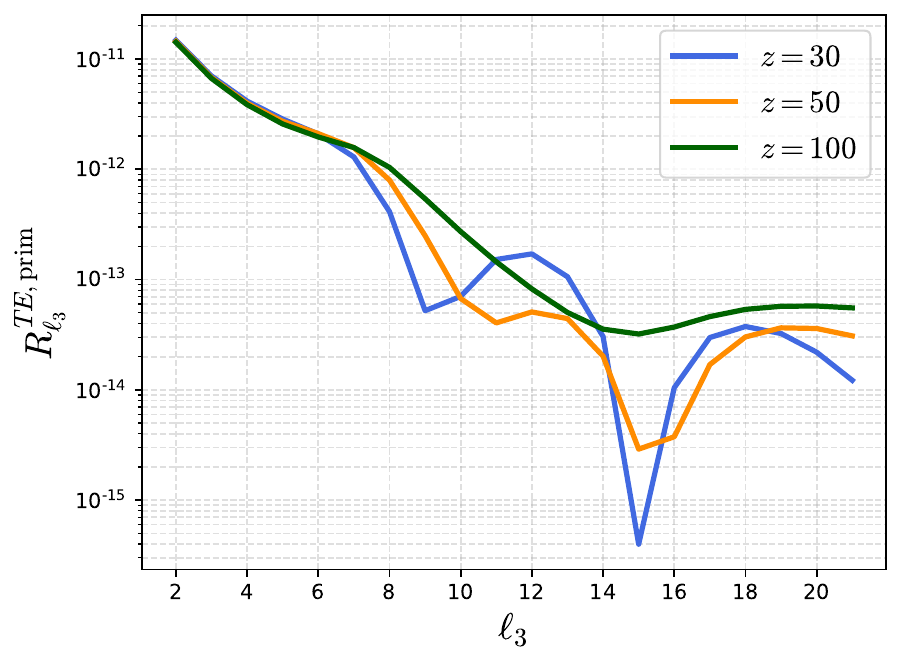} 
\caption{Same as Fig. \ref{fig:RTpriml3} but with the addition of the CMB polarization field, Eq. \eqref{eq:RTEpriml3}.} \label{fig:RTEpriml3}
\end{figure}

Next, we compute the Fisher information from cross-correlations with the addition of the CMB polarization ($E$-mode) field. Considering the vector of observables
\begin{equation}
\mathcal O_X = 
\begin{pmatrix}
	 B^{21-21-\rm T, \, prim}_{\ell_1 \ell_2 \ell_3}(z) \\
 B^{21-21-\rm E, \, prim}_{\ell_1 \ell_2 \ell_3}(z)
	\end{pmatrix},
\end{equation}
the Fisher matrix can be generalized
\begin{equation} \label{eq:Fisher_pol}
F(z) = f_{\rm sky} \, \sum_{\ell_3 \leq \ell_2 \leq \ell_1} \, \sum_{X Y} \, \mathcal O_X \, \Sigma^{-1}_{X Y} \,   \mathcal O_Y  \, ,
\end{equation}
where we introduced the following inverse covariance matrix 
\begin{equation} \label{eq:sigma}
\Sigma^{-1} = \frac{1}{\Delta_{\ell_1 \ell_2} \, C^{21}_{\ell_1}(z) \, C^{21}_{\ell_2}(z) } \, \begin{pmatrix}
	 \frac{C^{EE}_{\ell_3}}{\left(  C^{TT}_{\ell_3} C^{EE}_{\ell_3} - (C^{TE}_{\ell_3})^2 \right)} & \,\,\,\,  - \frac{C^{TE}_{\ell_3}}{\left(C^{TT}_{\ell_3} C^{EE}_{\ell_3} - (C^{TE}_{\ell_3})^2 \right)}  \\
                                          \\
	 - \frac{C^{TE}_{\ell_3}}{\left(C^{TT}_{\ell_3} C^{EE}_{\ell_3} - (C^{TE}_{\ell_3})^2 \right)} & \,\,\,\, \frac{C^{TT}_{\ell_3}}{\left(C^{TT}_{\ell_3} C^{EE}_{\ell_3} - (C^{TE}_{\ell_3})^2 \right)}
	\end{pmatrix} \, .
\end{equation}
In Fig. \ref{fig:forecastfNLTpE} we show the expected uncertainty in measuring $\fnl{loc}$ including the information coming from the CMB polarization field. We still have a saturation at $\ell_{\rm max} \approx 10$. The uncertainty per redshift slice improves by about a factor of $2$ with respect to the $T$-mode only case, with larger improvements at lower redshifts. Heuristically, the relevant contribution to the largest CMB $E$-mode anisotropies comes from the re-ionization epoch, which is closer in time to the lower redshift of the Dark Ages. This results in an opposite correction to the redshift increasing trend in the Fisher information of the $T$-only case and the redshift dependence of $\sigma(\fnl{loc})_{T+E}$ gets almost completely erased. A quantitative representation of this is displayed in Fig. \ref{fig:RTEpriml3}, where we plot the following $\ell_3$-dependence of each term in the Fisher matrix summation
\begin{align}  \label{eq:RTEpriml3}
R^{TE, \, \rm prim}_{\ell_3}(z) = \sum_{X, Y = T, E} \, \frac{ (-1)^x \, C^{XY}_{\ell_3}}{\left(  C^{TT}_{\ell_3} C^{EE}_{\ell_3} - (C^{TE}_{\ell_3})^2 \right)} \, & \left[\int \frac{d k_3}{k_3} \, j_{\ell_3}(k_3 r(z)) \, {\cal T}_{\ell_3(s)}^{X}(k_3) \,\mathcal A_s(k_3) \right]  \,  \nonumber \\
& \times \left[\int \frac{d k_3}{k_3} \, j_{\ell_3}(k_3 r(z)) \, {\cal T}_{\ell_3(s)}^{Y}(k_3) \,\mathcal A_s(k_3) \right] \, , 
\end{align}
with $x = 0$ when $X = Y$ and $x = 1$ when $X \neq Y$. As we can see, up to $\ell_3 \approx 7$ there is no appreciable redshift dependence. From the bottom panel of Fig. \ref{fig:forecastfNLTpE} we estimate the following scaling of $\fnl{loc}$ that could be reached with a single redshift measurement
\begin{equation} \label{eq:fNLTpE}
\sigma(\fnl{loc})_{T+E} \simeq \, 0.1  \, f^{-1/2}_{\rm sky}  \, \left(\frac{10^5}{\ell_{1, \rm max}}\right) \, .
\end{equation}


\subsection{Marginalization over secondaries}

As shown in Sec. \ref{sec:2121X}, besides the primordial signal we have to address secondary contributions, similar to those that appear for the 21-cm auto-bispectrum (e.g. \cite{Munoz:2015eqa}). Typically, when dealing with secondaries we can consider two options: we can either remove the expected secondaries from the data as it is done in the analysis of the CMB primary bispectrum (see e.g. \cite{Hanson:2009kg,Lewis:2011fk}), or marginalize the data over the secondary signal, as shown e.g. in Ref. \cite{Munoz:2015eqa} for the auto 21-cm bispectrum and trispectrum case. The best solution depends on the capacity to model the secondary signal with enough precision and on the relative strength of the secondary signal with respect to the primordial signal.

\begin{figure}  
        \includegraphics[width=\linewidth]{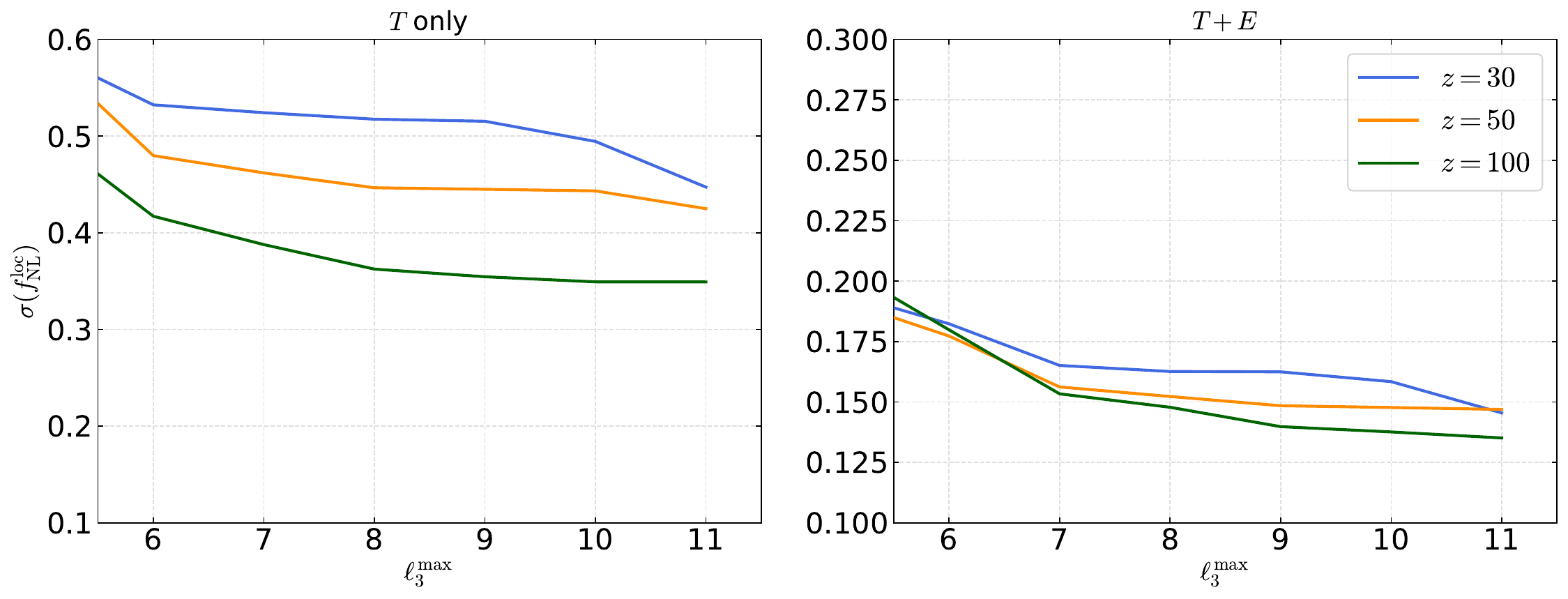} 
    \caption{Expected 1$\sigma$ uncertainty on $\fnl{loc}$ per single redshift slice after marginalization over secondaries. Angular resolution is the same adopted in Sub. \ref{sub:singlered}.} \label{fig:forecastfNL_marg}
\end{figure}

By including secondaries in the observed $\langle 21-21-\rm T \rangle$ bispectrum we have (we remind that the terms $\mathcal I^{\rm prim, 1}$ and $\mathcal I^{\rm sec, 1}$ largely dominate over the others)
\begin{equation} \label{eq:bispectrum_full}
B^{21-21-\rm T}_{\ell_1 \ell_2 \ell_3}  = \fnl{loc} \, \tilde B^{21-21-\rm T, \, prim}_{\ell_1 \ell_2 \ell_3}(z) + f_1(z) \, \tilde B^{21-21-\rm T, \, sec, 1}_{\ell_1 \ell_2 \ell_3}(z) + f_2(z) \, \tilde B^{21-21-\rm T, \, sec, 2}_{\ell_1 \ell_2 \ell_3}(z) \, ,
\end{equation}
where 
\begin{align}
f_1(z) &= \alpha_1(z) \, \Big[\alpha_2(z) d_0 +  \alpha_3(z) \Big] \, , \nonumber \\
f_2(z) &= \alpha_1(z) \, \alpha_2(z) \, . 
\end{align}  
We can redefine the bispectrum in Eq. \eqref{eq:bispectrum_full} as 
\begin{equation}
B^{21-21-\rm T}_{\ell_1 \ell_2 \ell_3}(z)  =  B^{21-21-\rm T, \, sec}_{\ell_1 \ell_2 \ell_3}(z) + \Delta f_0 \, \tilde B^{21-21-\rm T, \, 0}_{\ell_1 \ell_2 \ell_3} + \Delta f_1 \, \tilde B^{21-21-\rm T, \, 1}_{\ell_1 \ell_2 \ell_3}(z) + \Delta f_2 \, \tilde B^{21-21-\rm T, \, 2}_{\ell_1 \ell_2 \ell_3}(z) \, ,
\end{equation}
where $B^{21-21-\rm T, \, sec}$ is the total secondary bispectrum, $\Delta f_i$'s represent the unknown residuals to the coefficients $f_i$'s and for convenience of notation we defined 
\begin{align}
\Delta f_0 & \equiv \fnl{loc} \, , \nonumber \\
\tilde B^{21-21-\rm T, \, 0} & \equiv \tilde B^{21-21-\rm T, \, \rm prim} \, , \nonumber \\
\tilde B^{21-21-\rm T, \, 1} & \equiv \tilde B^{21-21-\rm T, \, \rm sec, 1} \, , \nonumber \\
\tilde B^{21-21-\rm T, \, 2} & \equiv \tilde B^{21-21-\rm T, \, sec, 2} \, . 
\end{align}
As shown e.g. in \cite{Munoz:2015eqa} for each $\Delta f_i$'s we can define the following MVNH estimator
\begin{equation} \label{eq:estimatorfNL_marg}
\Delta  \hat{f_i}|_z =  \sum_j \sum_{\ell_3 \leq \ell_2 \leq \ell_1} \, (F^{-1})_{ij} \frac{\left(B^{21-21-\rm T, \, obs}_{\ell_1 \ell_2 \ell_3}(z)-B^{21-21-\rm T, \, sec}_{\ell_1 \ell_2 \ell_3}(z)\right) \,\,\, \tilde B^{21-21-{\rm T}, \,j}_{\ell_1 \ell_2 \ell_3}(z)}{C^{21}_{\ell_1}(z) \, C^{21}_{\ell_2}(z) \, C^{\rm T}_{\ell_3}} \, ,
\end{equation}
where the Fisher matrix reads
\begin{equation} \label{eq:Fisher_marg}
F_{ij}(z) = f_{\rm sky} \, \sum_{\ell_3 \leq \ell_2 \leq \ell_1} \, \frac{\tilde B^{21-21-{\rm T}, \, i}_{\ell_1 \ell_2 \ell_3}(z) \, \tilde B^{21-21-{\rm T}, \, j}_{\ell_1 \ell_2 \ell_3}(z)}{\Delta_{\ell_1 \ell_2} \, C^{21}_{\ell_1}(z) \, C^{21}_{\ell_2}(z) \, C^{\rm T}_{\ell_3}} \, .
\end{equation}
The variances on the $\Delta f_i$'s are given by 
\begin{equation}
\sigma^2(\Delta f_i) =  (F^{-1})_{ii} \, .
\end{equation}
Similar expressions can be derived when we include the CMB $E$-mode polarization. In Fig. \ref{fig:forecastfNL_marg} we show the expected $1\sigma$ uncertainty in measuring $\fnl{loc}$ adopting the same angular resolution of the previous subsection after we marginalize over the secondary contributions. We see that ultimate detection prospects degrade less than a factor 3 in the $T$ only case and less than a factor 2 when adding CMB polarization. 

\subsection{Bias due to secondaries}

If we do not subtract or marginalize all the signal coming from the secondary contributions, it will introduce the following bias in the measurement of $\fnl{loc}$ 
\begin{equation} \label{eq:biasfNl}
\Delta_b \fnl{loc}(z) = n\% \,\, \times \frac{F_{\rm prim \, \times \, sec}(z)}{F_{\rm prim}(z)} \, ,
\end{equation}
where 
\begin{equation} \label{eq:Fisher_bias}
F_{\rm prim \, \times \, sec}(z) = f_{\rm sky} \, \sum_{\ell_3 \leq \ell_2 \leq \ell_1} \, \frac{\tilde B^{21-21-\rm T, \, prim}_{\ell_1 \ell_2 \ell_3}(z) \, B^{21-21-\rm T, \, \rm sec}_{\ell_1 \ell_2 \ell_3}(z)}{\Delta_{\ell_1 \ell_2} \, C^{21}_{\ell_1}(z) \, C^{21}_{\ell_2}(z) \, C^{\rm T}_{\ell_3}} \, ,
\end{equation}
$F_{\rm prim}(z)$ is as in Eq. \eqref{eq:Fisher}, and $n\%$ is the percentage of total secondary signal that still remains after we remove or marginalize it from the data. The generalization to the case where we add the CMB polarization field is straightforward. 

\begin{figure}[h!]
        \centering
        \includegraphics[width=\linewidth]{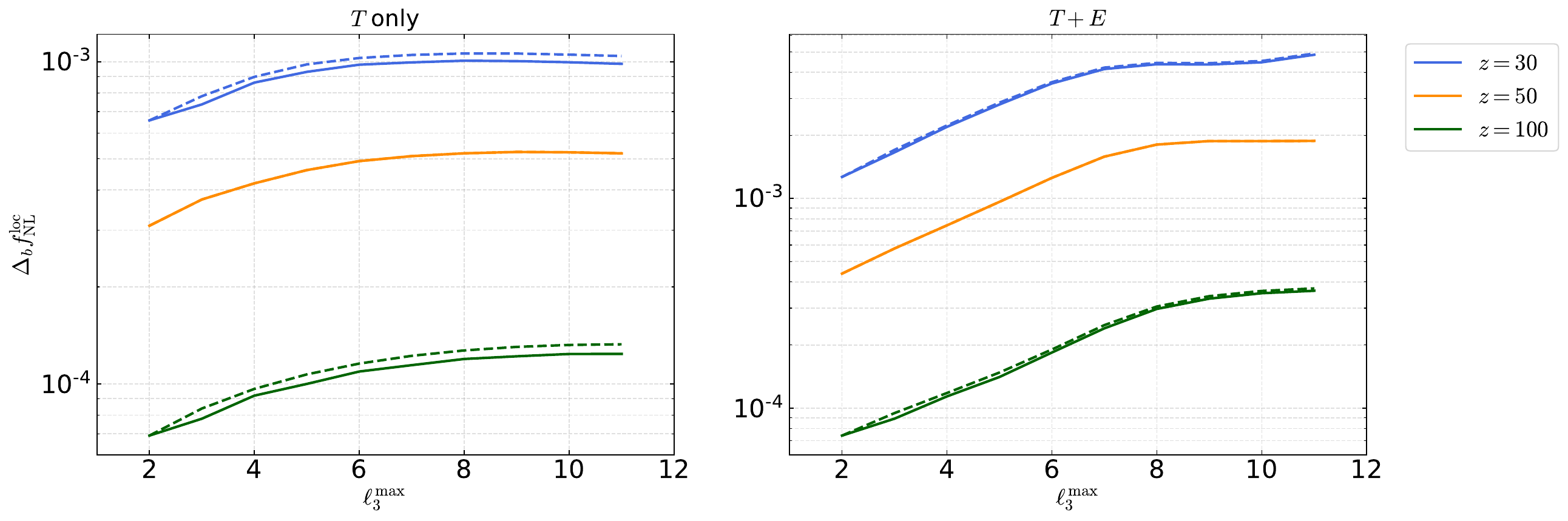} 
\caption{Expected bias in the measurement of $\fnl{loc}$ coming from $1\%$-residual secondaries after subtraction. Dashed lines are the same quantities obtained by neglecting the $\mathcal I^{\rm sec, 2}$ contribution.} \label{fig:biasfNLvsl3}
\end{figure}

Here we assume that the coefficients $f_i$'s that describe the secondary signal are known with a reasonable percent-level error, resulting in $n\% \simeq 1\%$. By employing Eq. \eqref{eq:biasfNl}, in Fig. \ref{fig:biasfNLvsl3} we show the resulting bias after we have removed the secondary signal from the data.  From the figure we understand that the $\mathcal I^{\rm sec, 2}$ term can in principle be neglected. We note that when only $T$ is considered, the bias on the measurement of $\fnl{loc}$ does not have a strong dependence on $\ell^{\rm max}_3$. In addition, the bias decreases with redshift. In contrary, when polarization is added, there is an appreciable growing bias with $\ell^{\rm max}_3$. 

\begin{figure}[h!]
        \centering
        \includegraphics[width=\linewidth]{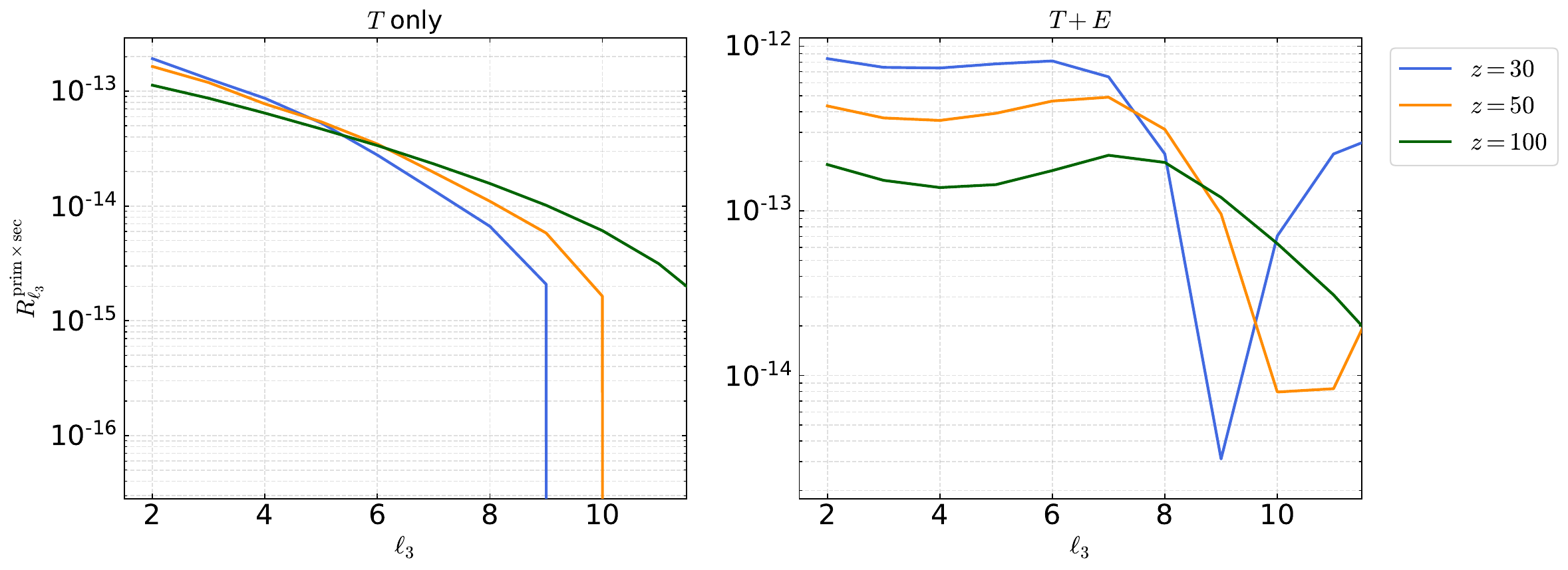} 
\caption{$\ell_3$ dependence of each term in the $F_{\rm prim \, \times \, sec}$ Fisher matrix, Eqs. \eqref{eq:RTsecl3} and \eqref{eq:RTEsecl3} .} \label{fig:Rsec}
\end{figure}

We explain this by introducing the equivalent of Eqs. \eqref{eq:RTpriml3} and \eqref{eq:RTEpriml3} but for the $(\rm prim \, \times \, sec)$ Fisher matrix, i.e., 
\begin{equation}  \label{eq:RTsecl3}
R^{T, \, \rm prim \, \times \, sec}_{\ell_3}(z) = \frac{\left[\int \frac{d k_3}{k_3} \, j_{\ell_3}(k_3 r(z)) \, {\cal T}_{\ell_3(s)}^{T}(k_3) \, \mathcal A_s(k_3) \right] \, \left[\int \frac{d k_3}{k_3} \, j_{\ell_3}(k_3 r(z)) \, {\cal T}_{\ell_3(s)}^{T}(k_3) \, \mathcal M_b\left(k_3\right) \, \mathcal A_s(k_3) \right]}{C^{\rm T}_{\ell_3}}  \, ,
\end{equation} 
and
\begin{align}  \label{eq:RTEsecl3}
R^{T+E, \, \rm prim \, \times \, sec}_{\ell_3}(z) = \sum_{X, Y = T, E} \, \frac{ (-1)^x \, C^{XY}_{\ell_3}}{\left(  C^{TT}_{\ell_3} C^{EE}_{\ell_3} - (C^{TE}_{\ell_3})^2 \right)} \, & \left[\int \frac{d k_3}{k_3} \, j_{\ell_3}(k_3 r(z)) \, {\cal T}_{\ell_3(s)}^{X}(k_3) \, \mathcal A_s(k_3) \right]  \,  \nonumber \\
& \times \left[\int \frac{d k_3}{k_3} \, j_{\ell_3}(k_3 r(z)) \, {\cal T}_{\ell_3(s)}^{Y}(k_3) \, \mathcal M_b\left(k_3\right) \, \mathcal A_s(k_3) \right] \, . 
\end{align}
These give the $\ell_3$ scaling of each term in the $(\rm prim \, \times \, sec)$ Fisher matrix (recall that contributions from $\mathcal I^{\rm sec, 2}$ can be neglected). We show these quantities in Fig. \ref{fig:Rsec}. The linear dependence of these terms on the baryon matter transfer function (which decreases with the redshift, see Fig. \ref{fig:est}) explains why the bias induced by secondaries generally decreases with redshift. Another thing to point out is the $\ell_3$-dependence of these $R_{\ell_3}$'s. As can be seen from the left panel of Fig. \ref{fig:Rsec}, in the $T$ case there is a modest decrease in $R_{\ell_3}$, which at a certain point turns into a rapid decrease. This explains the relatively slow increase and eventual saturation of the bias in the left panel of Fig. \ref{fig:biasfNLvsl3}. From the right panel of Fig. \ref{fig:Rsec} we observe that for $T+E$, $R_{\ell_3}$ remains almost constant for the first $\ell_3$ multipoles, leading to an important growth of the bias with $\ell^{\rm max}_3$ (right panel of Fig. \ref{fig:biasfNLvsl3}). 

Based on Fig. \ref{fig:biasfNLvsl3} we expect the following biases when considering the information coming from squeezed triangular configurations with $\ell_3 <11$ with a percent-level uncertainty on the determination of the coefficients $f_1$ and $f_2$:
\begin{align}
\Delta_{b} \fnl{loc}(z=30)_T \simeq & 10^{-3} \, , \qquad \qquad \qquad \, \Delta_{b} \fnl{loc}(z=30)_{T+E} \simeq 4.5 \times  10^{-3} \, , \nonumber \\
\Delta_{b} \fnl{loc}(z=50)_T \simeq& 5 \times 10^{-4}  \, , \qquad \qquad \,\,\,\Delta_{b} \fnl{loc}(z=50)_{T+E} \simeq  1.8 \times 10^{-3} \, ,  \nonumber \\
\Delta_{b} \fnl{loc}(z=100)_T \simeq & 1.2 \times 10^{-4}  \, , \qquad \qquad \Delta_{b} \fnl{loc}(z=100)_{T+E} \simeq  3.5 \times 10^{-4} \, . 
\end{align}
These biases are more than one order of magnitude smaller than the forecasted constraints on $\fnl{loc}$ reachable with a single redshift measurement of the 21-cm field anisotropies, which is of the order $\fnl{loc} \sim 10^{-1}$ for $\ell_{1}^{\rm max} = 10^5$ as shown in Eqs. \eqref{eq:fNLT} and \eqref{eq:fNLTpE}. Therefore, under the experimental setup considered in this work, if secondaries are modeled with a reasonable uncertainty, the resulting bias after simple subtraction from the data does not spoil projected constraints.  

\subsection{Tomography}

\begin{figure}    
        \centering
        \includegraphics[width=12cm]{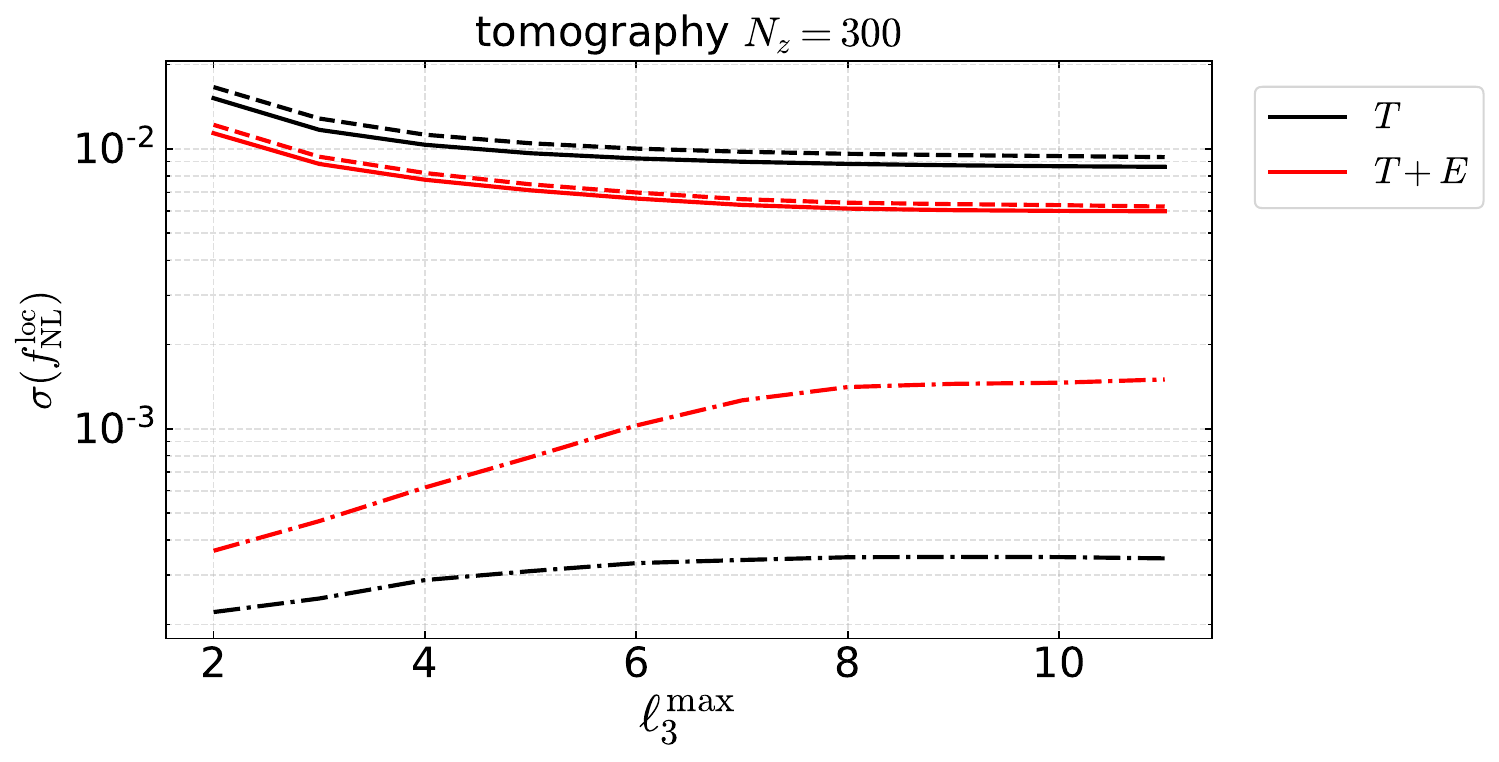} 

    \caption{Expected 1$\sigma$ uncertainty combining information from $300$ uncorrelated redshift slices from $z=30$ to $z=100$. Solid lines: with no secondaries marginalized. Dashed lines: with secondaries marginalized. Dot-dashed lines: expected biases in the measurement of $\fnl{loc}$ coming from $1\%$-residual secondaries after subtraction.} \label{fig:forecast_tom}
\end{figure}

As pointed out in previous literature (e.g. Refs. \cite{Pillepich:2006fj,Munoz:2015eqa}), one of the advantages of the 21-cm anisotropies is that we can combine the information from different redshift slices. As shown e.g. in Ref. \cite{Munoz:2015eqa}, if we observe the 21-cm fluctuations on very small scales ($\ell_1>10^4$), the correlation length $\xi_\nu$ would be $< 0.1$ MHz. This means that if we observe the 21-cm fluctuations between 14 MHz ($z = 100$) and 45 MHz ($z = 30$) with a frequency resolution $\Delta \nu \geq 0.1$, each separate redshift slice will be uncorrelated. The resulting number of uncorrelated redshift slices reads  
\begin{equation}
N_z \simeq 30 \, \frac{1 \, \mbox{MHz}}{\Delta \nu} \, .
\end{equation}
We can accumulate information from each redshift slice and sum-up this information in the following total Fisher matrix 
\begin{equation}
F_{ij}^{\rm tom} = \sum_z F_{ij}(z) \, .
\end{equation}
Using the results of the previous subsection and assuming an experiment with frequency resolution $\Delta \nu\geq 0.1$ (so $N_z = 300$) in Fig. \ref{fig:forecast_tom} we show the expected minimum detectable value of $\fnl{loc}$, together with the bias to its measurement that will occur due $1\%$ residual (unsubtracted) secondaries. We see that marginalizing over secondaries does not significantly affect ultimate constraints. From this plot we get the following projected constraints
\begin{align} \label{eq:deltafNLtom}
&\sigma(\fnl{loc})^{\rm tom}_{T} \simeq \, 9 \times 10^{-3} \, f^{-1/2}_{\rm sky} \, \left(\frac{\Delta \nu}{0.1 \, \mbox{MHz}}\right)^{1/2}  \,  \left(\frac{10^5}{\ell^{\rm max}_1}\right)^2  \qquad \qquad (\Delta_{b} \fnl{loc})^{\rm tom}_T \simeq  3.5 \times 10^{-4} \, , \\
&\sigma(\fnl{loc})^{\rm tom}_{T+E} \simeq \, 6 \times 10^{-3} \, f^{-1/2}_{\rm sky} \, \left(\frac{\Delta \nu}{0.1 \, \mbox{MHz}}\right)^{1/2}  \, \left(\frac{10^5}{\ell^{\rm max}_1}\right)^2 \qquad \qquad (\Delta_{b} \fnl{loc})^{\rm tom}_{T+E} \simeq  1.4 \times 10^{-3} \,  .  \label{eq:deltafNLtom2}
\end{align}
Here the residual biases are still lower than the expected constraints for $\ell_{1}^{\rm max} = 10^5$ and $\Delta \nu = 0.1 \, \mbox{MHz}$, suggesting that the ultimate constraints are not spoiled by subtracting secondaries, providing that the latters are modeled with reasonable precision.        

\subsection{Non-Gaussian covariance}

When computing the cosmic variance-limited covariance matrix of the $\langle \rm 21-21-CMB \rangle$ cross-correlations we assumed that Gaussian terms dominate over non-Gaussian ones, resulting in a diagonal covariance matrix. In reality, (see e.g. \cite{Floss:2022wkq}), secondary non-Gaussianity will induce off-diagonal terms in the covariance matrix, resulting in a reduction of the information content and hence a worse constraint on $\fnl{loc}$ (or any type of NG). For example accounting for non-Gaussian terms the Fisher matrix of Eq. \eqref{eq:Fisher} can be generalized as
\begin{equation} \label{eq:Fisher_full}
F(z) = f_{\rm sky} \, \sum_{L} \sum_{L'} \, \tilde B^{21-21-\rm T, \, prim}_{L}(z) \,\, C^{-1}_{L L'} \,\, \tilde B^{21-21-\rm T, \, prim}_{L'}(z) \, ,
\end{equation}
where $L=(\ell_1, \ell_2, \ell_3)$ stands for a particular multipole triangular configuration where $\ell_3 \leq \ell_2 \leq \ell_1$. Here the full Wick-expanded covariance matrix contains $4$ terms, one given by the product of two 21-cm power spectra and one CMB power spectrum, one given by the product of two cross-correlated bispectra, one given by the product of a CMB power spectrum and the 21-cm trispectrum, and the last one given by the fully connected cross-correlated six-point function, the pentaspectrum. We can write this decomposition schematically as
\begin{align}
C_{L L'} = \delta_{L L'} \, & C^{\rm CMB}_{L L'} C^{21}_{L L'} C^{21}_{L L'} \nonumber \\
&+ B_L^{\rm 21-21-CMB} \, B_{L'}^{\rm 21-21-CMB} + T_{L L'}^{\rm 21-21-21-21} \, C^{\rm CMB}_{L L'} \nonumber\\
&+ P^{\rm 21-21-21-21-CMB-CMB}_{L L'} \, .
\end{align}
The first term of this equation is the diagonal covariance matrix employed in Eq.~\eqref{eq:Fisher}. As shown in previous literature (see e.g. \cite{Biagetti:2021tua}) the $B \times B$ and $T \times C$ terms can give significant contributions when summing over squeezed triangular configurations that share the same long mode (with $T \times C \simeq B \times B$), while the pentaspectrum can in principle be neglected. Using the results of subsections \ref{sub:prim} and \ref{sub:sec} we can estimate the strength of the $B \times B$ term over the diagonal (Gaussian) term. We start by considering the primordial contributions of Eq. \eqref{eq:cross_full_prim_tot} and taking the temperature mode. We find
\begin{align}
\frac{B_L^{\rm 21-21-T, \, prim} \, B_{L'}^{\rm 21-21-T, \, prim} }{\Delta_{\ell_1 \ell_2} \, C^{\rm T}_{L L'} C^{21}_{L L'} C^{21}_{L L'} \delta_{L L'}}\simeq & 180 \, (\fnl{loc})^2 \, \sqrt{\ell_1 \, \ell_1'} \, \left(\frac{\ell_1}{\ell'_1}\right)^3 \, \left(\frac{\mathcal{M}_b(\ell'_1/r)}{\mathcal{M}_b(\ell_1/r)}\right)^2 \, \frac{1}{\Delta_{\ell_1 \ell_2}} \nonumber \\ 
 \qquad \qquad \qquad &\times \frac{\left[\int \frac{d k_3}{k_3} \, j_{\ell_3}(k_3 r(z)) \, {\cal T}_{\ell_3(s)}^{T}(k_3) \,\mathcal A_s(k_3) \right] \, \left[\int \frac{d k_3}{k_3} \, j_{\ell'_3}(k_3 r(z)) \, {\cal T}_{\ell'_3(s)}^{T}(k_3) \,\mathcal A_s(k_3) \right]}{C^{\rm T}_{\ell_3}} \, . 
\end{align}
The configurations for which this term is maximized are those where $\ell_3= \ell'_3 =2$, with the $\ell_3$ dependent term becoming equal to $R^{T, \, \rm prim}_{\ell_3}$, Eq. \eqref{eq:RTpriml3}. Therefore, we have 
\begin{align}
\frac{B_L^{\rm 21-21-T, \, prim} \, B_{L'}^{\rm 21-21-T, \, prim}}{\Delta_{\ell_1 \ell_2} \, C^{\rm T}_{L L'} C^{21}_{L L'} C^{21}_{L L'} \delta_{L L'}} < & 1.8 \times 10^{-9} \, (\fnl{loc})^2 \, \sqrt{\ell_1 \, \ell_1'} \, \left(\frac{\ell_1}{\ell'_1}\right)^3 \, \left(\frac{\mathcal{M}_b(\ell'_1/r)}{\mathcal{M}_b(\ell_1/r)}\right)^2 \, \frac{1}{\Delta_{\ell_1 \ell_2}} \, . 
\end{align}
This term is enhanced when $\ell'_1<\ell_1$. The maximum enhancement arises when $\ell'_1 = 10^4$ and $\ell_1 = 10^5$, yielding an upper bound
\begin{align}
\frac{B_L^{\rm 21-21-T, \, prim} \, B_{L'}^{\rm 21-21-T, \, prim}}{\Delta_{\ell_1 \ell_2} \, C^{\rm T}_{L L'} C^{21}_{L L'} C^{21}_{L L'} \delta_{L L'}} < & 7 \times 10^{-3} \, (\fnl{loc})^2  \, . 
\end{align}
As we are interested in probing  $\fnl{loc} <1$, it follows that off-diagonal terms coming from small primordial non-Gaussianities can be neglected (this estimation can be extended to the $E$-mode case as well). What about the contribution from secondaries? As we have seen above, secondary contributions are dominated by the term $\mathcal I^{\rm sec, 1}$, Eq. \eqref{eq:I1limberm}, with a shape similar to the primordial signal. In analogy we can obtain upper bounds
\begin{align}
\frac{B_L^{\rm 21-21-T, \, sec} \, B_{L'}^{\rm 21-21-T, \, sec}}{\Delta_{\ell_1 \ell_2} \, C^{\rm T}_{L L'} C^{21}_{L L'} C^{21}_{L L'} \delta_{L L'}} <  7 \times  10^{-5} \, ,  \qquad \qquad \frac{B_L^{\rm 21-21-T, \, prim} \, B_{L'}^{\rm 21-21-T, \, sec}}{\Delta_{\ell_1 \ell_2} \, C^{\rm T}_{L L'} C^{21}_{L L'} C^{21}_{L L'} \delta_{L L'}} <  7 \times 10^{-4} \, \fnl{loc} \, . 
\end{align}
From these upper bounds it follows that the covariance matrix of squeezed triangular configurations (which are those that provide the relevant Fisher information) is dominated by the diagonal, i.e. we do not expect significant non-Gaussian off-diagonal contribution to our $\fnl{loc}$ forecast.

\section{Discussion and Conclusion} \label{sec:concl}

In this paper we have investigated cross-correlations between 21-cm anisotropies coming from the Dark Ages and CMB $T$- and $E$-mode anisotropies, which are generated by the same primordial seed. Since CMB anisotropies are generated at scales that are generally much larger than the 21-cm brightness temperature anisotropies, the cross-correlated two-point functions ($C_{\ell}^{\rm 21-CMB}$) are basically vanishing due to momenta conservation. Conversely, the three-point cross-correlations between the 21-cm and CMB signals ($B_{\ell_1, \,\ell_2, \,\ell_3}^{\rm 21-21-CMB}$) can be non-zero by momenta conservation and allow for constraining highly squeezed triangle configurations. Our final results target a 21-cm anisotropy detection corresponding to the scales that will be reachable by experiments on the moon ($1 \, \mbox{Mpc}^{-1}<k < 10 \, \mbox{Mpc}^{-1}$, or on multipoles $\ell_1 \simeq 10^4-10^5$). In this analysis we considered both the primordial signal from local non-Gaussianity and the secondary signal which is introduced by the non-linear evolution of the 21-cm field. 

Let us summarize our main findings. First, contrary to the 21-cm auto bispectrum, we found that secondaries introduce only a small bias (in the worst case scenario $\Delta_{b} {\fnl{loc}}|_{T+E} \lesssim 10^{-3}$ assuming that secondaries can be modeled and removed with percent-level residuals) to the measurement of $\fnl{loc}$. This is due to the secondary signal being damped by the very small baryonic transfer function on the largest CMB scales ($10^{-4} \, \mbox{Mpc}^{-1}<k < 10^{-3} \, \mbox{Mpc}^{-1}$, or $\ell_3 \lesssim 10$), which dominate the Fisher information. We have shown that with marginalization techniques we can completely remove this bias, similarly to the 21-cm auto bispectrum analyses, without significantly affecting projected constraints.

Second, we found that by combining $T+E$ anisotropies, $\fnl{loc} \sim 10^{-1}$ is in principle reachable using only a single redshift measurement of 21-cm from the Dark Ages. Moreover, combining measurements of different uncorrelated redshift slices we can push this bound down to $\fnl{loc} \sim 6 \times 10^{-3}$. This constraint involves ultra-squeezed triangular configurations that are complementary to those targeted by either the CMB and 21-cm auto bispectra.

Third, we have estimated that the effect of off-diagonal terms in the covariance do not significantly effect the projected constraints. We stress that this particular characteristic of the $\langle 21-21- \rm CMB \rangle $ cross-correlation would make it more competitive than the 21-cm auto bispectrum in the ultimate search for local non-Gaussianities. For the auto bispectrum, single redshift constraints on $\fnl{loc}$ are subject to a damping that in the best case scenario ($z = 100$) is of at least one order of magnitude (see Fig. 1 of Ref.~\cite{Floss:2022wkq}). If not accounted for, this would limit the search for pnG’s using the 21-cm auto bispectrum by more than one order of magnitude.

Forth, in our results we have completely neglected the effect of velocity terms on the 21-cm field. This assumption is well motivated given the frequency and angular resolutions selected in this work as hypothetical experiment. However, given the fact that the final Fisher matrix estimated in Eq. \eqref{eq:Fisher_approx} is mostly sensitive to the details of the CMB field and the angular resolution of a given 21-cm experiment, we argue that our results can be qualitatively applied for a generic frequency resolution $\Delta \nu$. Of course, a detailed analysis including velocity effects will provide more precise forecasts in the regime $\Delta \nu < 0.1$ MHz, while likely resulting in the same order of magnitude results found in this paper. 

Finally, we point out that we have assumed that a cosmic-variance limited detection of the 21-cm brightness temperature from the Dark Ages is possible. However, a realistic detection will be limited both by the experimental noise and severe (extra-galactic) foreground contamination. Previous works (see e.g. \cite{Cole:2019zhu} and Refs. therein) have shown that the level of expected noise of lunar experiments should allow one to measure the expected primordial 21-cm signal for the range of scales proposed in this work. However, the largest contamination to the signal is represented by foregrounds which are expected to be several order of magnitude larger than the primordial signal (see e.g. \cite{Burns:2021ndk}). We point out that different methods have been proposed so far to remove such a contamination. The most promising so far relies on the fact that when decomposing a given Fourier space signal in the $(k_\parallel, k_\perp)$ components, the so-called foreground ‘wedge’ (see e.g. \cite{Pober:2013jna,Pober:2014lva})
\begin{equation}
k_\parallel < k_{\rm wedge} \, k_\perp
\end{equation}
is less affected (or relatively clean) by foreground contamination. While this will reduce the overall signal-to-noise ratio for detecting the 21-cm anisotropies, it represents a way to have a clean detection as assumed in the present analysis. We leave this and other suggested extensions for future work. 

\section{Acknowledgements}

G.O. and P.D.M. acknowledge support from the Netherlands organization for scientific research (NWO) VIDI grant (dossier 639.042.730). G.O. thanks the National Energy Research Scientific Computing Center for providing access to the Perlmutter computing cluster. T.F. is supported by the Fundamentals of the Universe research program within the University of Groningen and thanks the Center for Information Technology of the University of Groningen for providing access to the Hábrók high performance computing cluster.

\appendix

\section{Computations} \label{app:computations}

In this appendix we show some explicit computational steps to get the primordial and secondary contributions to the $\langle \rm 21-21-CMB \rangle$ bispectra. 

\subsection{Primordial} \label{app:computations_prim}

We compute the primordial contribution to the $\langle \rm 21-21-CMB \rangle$  cross-correlations by inserting Eq. \eqref{eq:bisp_prim_scal} into Eq. \eqref{eq:cross_full_exp}. We find
\begin{align} \label{eq:cross_full_prim}
\langle a_{\ell_1 m_1}^{21} a_{\ell_2 m_2}^{21} a^X_{\ell_3 m_3}\rangle = & \frac{8}{\pi^3} \, i^{\ell_1 + \ell_2 + \ell_3} \, \int dr' \, \int dr'' \, W_{r(z)}(r') \, W_{r(z)}(r'') \, \int_0^\infty \, y^2 \, dy \, \left[\prod_{i=1}^3 \int d^3 k_i \, Y_{\ell_i m_i}^*(\hat{k}_i)  \right] \, \nonumber \\
& \quad \times \left[\prod_{n=1}^3 \sum_{L_n M_n} \, (-1)^{L_n/2} \,  j_{L_n}(k_i y) \,Y_{L_n M_n}^*(\hat{k}_i) \right] \, h_{L_1 L_2 L_3}^{0 0 0} \, \begin{pmatrix}
	L_1 & L_2 & L_3 \\
	M_1 & M_2 & M_3
	\end{pmatrix}  \nonumber \\
& \quad \times j_{\ell_1}(k_1 r') \, j_{\ell_2}(k_2 r'') \, {\cal T}_{\ell_3(s)}^{X}(k_3) \, \times \left(\alpha_1(r') \, \alpha_1(r'') \,  \mathcal{M}_b(k_{1}, r') \,  \mathcal{M}_b(k_{2}, r'') \right) \nonumber \\
& \quad \times B^{\rm loc}_{\zeta\zeta\zeta}(k_{1},k_{2},k_{3}) \, .
\end{align} 
By reordering the radial and angular integrations, we obtain
\begin{align} \label{eq:cross_full_prim2}
\langle a_{\ell_1 m_1}^{21} a_{\ell_2 m_2}^{21} a^X_{\ell_3 m_3}\rangle = & \frac{8}{\pi^3} \, i^{\ell_1 + \ell_2 + \ell_3} \, \int dr' \, \int dr'' \, W_{r(z)}(r') \, W_{r(z)}(r'') \, \int_0^\infty \, y^2 \, dy \, \left[\prod_{i=1}^3 \int d k_i \, k_i^2 \, \right] \, \nonumber \\
& \quad \times \left[\prod_{n=1}^3 \sum_{L_n M_n} \, (-1)^{L_n/2} \,  j_{L_n}(k_i y) \right] \, h_{L_1 L_2 L_3}^{0 0 0} \, \begin{pmatrix}
	L_1 & L_2 & L_3 \\
	M_1 & M_2 & M_3
	\end{pmatrix}  \nonumber \\
& \quad \times j_{\ell_1}(k_1 r') \, j_{\ell_2}(k_2 r'') \, {\cal T}_{\ell_3(s)}^{X}(k_3) \, \times \,  \mathcal{M}_b(k_{1}, r') \,  \mathcal{M}_b(k_{2}, r'') \times \, B^{\rm loc}_{\zeta\zeta\zeta}(k_{1},k_{2},k_{3})  \nonumber \\
& \quad \times \int d \hat k_1 \, Y_{\ell_1 m_1}^*(\hat{k}_1) \, Y_{L_1 M_1}^*(\hat{k}_1) \int d \hat k_2 \, Y_{\ell_2 m_2}^*(\hat{k}_2) \, Y_{L_2 M_2}^*(\hat{k}_2) \nonumber \\
& \quad \times  \int d \hat k_3 \, Y_{\ell_3 m_3}^*(\hat{k}_3) \, Y_{L_3 M_3}^*(\hat{k}_3) \times \left(\alpha_1(r') \, \alpha_1(r'') \, \right)  \, .
\end{align} 
From now on we will remove the radial $r$ dependencies for simplicity of notation.  

The angular integrations over $\hat{k}_i$ can be performed in terms of Wigner symbols using Eq. \eqref{eq:Gaunt_integral} of App. \ref{app:Wigner}. We get 
\begin{align} \label{eq:cross_full_prim3}
\langle a_{\ell_1 m_1}^{21} a_{\ell_2 m_2}^{21} a^X_{\ell_3 m_3}\rangle = & \frac{8}{\pi^3} \, (-1)^{\ell_1 + \ell_2 + \ell_3} \, h_{\ell_1 \ell_2 \ell_3}^{0 0 0} \, \begin{pmatrix}
	\ell_1 & \ell_2 & \ell_3 \\
	m_1 & m_2 & m_3
	\end{pmatrix}  \, \int_0^\infty \, y^2 \, dy \, \int d k_1 \, d k_2  \, d k_3  \, k_1^2 \, k_2^2 \, k_3^2    \nonumber \\
& \qquad \times j_{\ell_1}(k_1 y) \, j_{\ell_2}(k_2 y) \, j_{\ell_3}(k_3 y) \, \tilde \beta_{\ell_1}(k_1) \, \tilde \beta_{\ell_2}(k_2)  \,   {\cal T}_{\ell_3(s)}^{X}(k_3) \times \, B^{\rm loc}_{\zeta\zeta\zeta}(k_{1},k_{2},k_{3}) \, ,
\end{align} 
where we have defined 
\begin{equation} \label{eq:beta_bisp}
\tilde \beta_{\ell}(k) = \, \int dr \,  W_{r(z)}(r) \, \alpha_1(r) \, \mathcal{M}_b(k, r) \,  j_{\ell}(k r)  \, .
\end{equation}
By explicitly substituting the local ansatz \eqref{eq:BLocal} into Eq. \eqref{eq:cross_full_prim3} we get
\begin{align} \label{eq:cross_full_prim4}
\langle a_{\ell_1 m_1}^{21} a_{\ell_2 m_2}^{21} a^X_{\ell_3 m_3}\rangle = & \, \fnl{loc} \, \frac{192 \pi}{5} \, (-1)^{\ell_1 + \ell_2 + \ell_3} \, h_{\ell_1 \ell_2 \ell_3}^{0 0 0} \, \begin{pmatrix}
	\ell_1 & \ell_2 & \ell_3 \\
	m_1 & m_2 & m_3
	\end{pmatrix}  \, \int_0^\infty \, y^2 \, dy \, \int d k_1 \, d k_2  \, d k_3     \nonumber \\
& \qquad \times j_{\ell_1}(k_1 y) \, j_{\ell_2}(k_2 y) \, j_{\ell_3}(k_3 y) \, \tilde \beta_{\ell_1}(k_1) \, \tilde \beta_{\ell_2}(k_2) \, {\cal T}_{\ell_3(s)}^{X}(k_3) \, \nonumber  \\
& \qquad \times \, \left( \frac{k_3^2}{k_1 k_2}  \, \mathcal A_s(k_1) \mathcal A_s(k_2)  +  \frac{k_1^2}{k_2 k_3}  \, \mathcal A_s(k_2) \mathcal A_s(k_3) +  \frac{k_2^2}{k_1 k_3}  \, \mathcal A_s(k_1) \mathcal A_s(k_3)   \right) \, .
\end{align} 
By matching this last equation with Eq. \eqref{eq:decisot}, we read off the final angular averaged primordial contribution
\begin{align} \label{eq:cross_full_prim5}
B^{21-21-\rm X, \, prim}_{\ell_1 \ell_2 \ell_3} = & \, \fnl{loc} \, \frac{192 \pi}{5} \, (-1)^{\ell_1 + \ell_2 + \ell_3} \, \sqrt{\frac{(2 \ell_1+1) (2 \ell_2+1) (2 \ell_3+1)}{(4 \pi)}} \, \begin{pmatrix}
	\ell_1 & \ell_2 & \ell_3 \\
	0 & 0 & 0
	\end{pmatrix}   \, \int_0^\infty \, y^2 \, dy \, \int d k_1 \, d k_2  \, d k_3     \nonumber \\
& \qquad \times j_{\ell_1}(k_1 y) \, j_{\ell_2}(k_2 y) \, j_{\ell_3}(k_3 y) \, \tilde \beta_{\ell_1}(k_1) \, \tilde \beta_{\ell_2}(k_2) \, {\cal T}_{\ell_3(s)}^{X}(k_3) \nonumber  \\
& \qquad \times \, \left( \frac{k_3^2}{k_1 k_2}  \, \mathcal A_s(k_1) \mathcal A_s(k_2)  +  \frac{k_1^2}{k_2 k_3}  \, \mathcal A_s(k_2) \mathcal A_s(k_3)  +  \frac{k_2^2}{k_1 k_3}  \, \mathcal A_s(k_1) \mathcal A_s(k_3)  \right) \, .
\end{align}
Notice that it is convenient to separate the momenta integration as follows
\begin{align} \label{eq:cross_full_prim6}
B^{21-21-\rm X, \, prim}_{\ell_1 \ell_2 \ell_3}(z) = & \, \fnl{loc} \, \frac{192 \pi}{5} \, (-1)^{\ell_1 + \ell_2 + \ell_3} \, \sqrt{\frac{(2 \ell_1+1) (2 \ell_2+1) (2 \ell_3+1)}{(4 \pi)}} \, \begin{pmatrix}
	\ell_1 & \ell_2 & \ell_3 \\
	0 & 0 & 0
	\end{pmatrix}   \nonumber \\
 & \qquad \qquad \times  \Big(\mathcal I^{\rm prim, 1}_{\ell_1 \ell_2 \ell_3} + \mathcal I^{\rm prim, 2}_{\ell_1 \ell_2 \ell_3} \Big) \, ,
\end{align}
where
\begin{align} 
\mathcal I^{\rm prim, 1}_{\ell_1 \ell_2 \ell_3} =  \int_0^\infty \, y^2 \, dy \left[\int \frac{d k_1}{k_1} \, j_{\ell_1}(k_1 y) \, \tilde \beta_{\ell_1}(k_1)  \,\mathcal A_s(k_1) \right] \nonumber  \\
\times \left[\int \frac{d k_2}{k_2} \, j_{\ell_2}(k_2 y) \, \tilde \beta_{\ell_2}(k_2) \,\mathcal A_s(k_2) \right] \nonumber  \\
\times \left[\int d k_3 \, k_3^2 \, j_{\ell_3}(k_3 y) \, {\cal T}_{\ell_3(s)}^{X}(k_3)  \right] \, , 
\end{align}
and 
\begin{align} 
\mathcal I^{\rm prim, 2}_{\ell_1 \ell_2 \ell_3} = \,\int_0^\infty \, y^2 \, dy \left[\int d k_1 \, k^2_1 \, j_{\ell_1}(k_1 y) \, \tilde \beta_{\ell_1}(k_1)   \right] \nonumber  \\
\times \left[\int \frac{d k_2}{k_2}\, j_{\ell_2}(k_2 y) \, \tilde \beta_{\ell_2}(k_2)  \,\mathcal A_s(k_2) \right] \nonumber \\
\times \left[\int \frac{d k_3}{k_3} \, j_{\ell_3}(k_3 y) \, {\cal T}_{\ell_3(s)}^{X}(k_3) \,\mathcal A_s(k_3) \right] \nonumber \\
+ (\ell_1 \leftrightarrow \ell_2) \, .
\end{align}
We can remove the radial dependence of the baryon transfer function and the $\alpha_i$ coefficients from the definition of the $\tilde \beta$ function in Eq. \eqref{eq:beta_bisp}, evaluating these parameters at the redshift $z$. We have explicitly checked that this provides negligible effects on the final numerical results as the baryon transfer functions and $\alpha_i$ coefficients are smooth functions of redshift. Therefore, we can re-write our final terms as
\begin{align}  \label{eq:I1prim}
\mathcal I^{\rm prim, 1}_{\ell_1 \ell_2 \ell_3} = \alpha^2_1(z) \, \int_0^\infty \, y^2 \, dy \left[\int \frac{d k_1}{k_1} \, j_{\ell_1}(k_1 y) \, \beta_{\ell_1}(k_1) \, \mathcal{M}_b(k_1, z) \,\mathcal A_s(k_1) \right] \nonumber  \\
\times \left[\int \frac{d k_2}{k_2} \, j_{\ell_2}(k_2 y) \,  \beta_{\ell_2}(k_2) \, \mathcal{M}_b(k_2, z) \, \mathcal A_s(k_2) \right] \nonumber  \\
\times \left[\int d k_3 \, k_3^2 \, j_{\ell_3}(k_3 y) \, {\cal T}_{\ell_3(s)}^{X}(k_3)  \right] \, , 
\end{align}
and 
\begin{align}  \label{eq:I2prim}
\mathcal I^{\rm prim, 2}_{\ell_1 \ell_2 \ell_3} = \alpha^2_1(z) \, \int_0^\infty \, y^2 \, dy \left[\int d k_1 \, k^2_1 \, j_{\ell_1}(k_1 y) \,  \beta_{\ell_1}(k_1) \, \mathcal{M}_b(k_1, z)  \right] \nonumber  \\
\times \left[\int \frac{d k_2}{k_2}\, j_{\ell_2}(k_2 y) \,  \beta_{\ell_2}(k_2) \, \mathcal{M}_b(k_2, z) \,\mathcal A_s(k_2) \right] \nonumber \\
\times \left[\int \frac{d k_3}{k_3} \, j_{\ell_3}(k_3 y) \, {\cal T}_{\ell_3(s)}^{X}(k_3) \,\mathcal A_s(k_3) \right] \nonumber \\
+ (\ell_1 \leftrightarrow \ell_2) \, .
\end{align}
with 
\begin{equation} \label{eq:beta_simpl}
\beta_{\ell}(k) = \, \int dr \,  W_{r(z)}(r)  \,  j_{\ell}(k r)  \, .
\end{equation}

\subsubsection*{Limber approximation}

Similarly to deriving Eq. \eqref{eq:21cmpower_limber}, we can simplify our results using the Limber approximation. By expanding the spherical harmonics describing the projection of the 21-cm field using Eq. \eqref{eq:Bessel_Limber}, the two terms above read
\begin{align} \label{eq:I1prim_Limber}
\mathcal I^{\rm prim, 1}_{\ell_1 \ell_2 \ell_3} =   \left(\frac{\pi}{2}\right)^2 \, \left(\frac{1}{\ell_2}\right)^3 \, \alpha^2_1(z)  \, \int_0^\infty  \,  dy \, y &\, \left[W_{r(z)}(y)\right]^2 \, \mathcal{M}_b(\ell_1/y)  \, \mathcal{M}_b(\ell_2/y) \, \mathcal A_s(\ell_2/y) \nonumber \\
&\times \left[\int \frac{d k_3}{k_3} \, j_{\ell_3}(k_3 y) \, {\cal T}_{\ell_3(s)}^{X}(k_3) \,\mathcal A_s(k_3) \right] \nonumber \\
&+ (\ell_1 \leftrightarrow \ell_2) \, ,
\end{align}
and
\begin{align} \label{eq:I2prim_Limber}
\mathcal I^{\rm prim, 2}_{\ell_1 \ell_2 \ell_3} =  \left(\frac{\pi}{2}\right)^2 \, \left(\frac{1}{\ell_1 \,\ell_2}\right)^3  \, \alpha^2_1(z)  \,\int_0^\infty \, dy \, y^4 &\, \left[W_{r(z)}(y)\right]^2 \, \mathcal{M}_b(\ell_1/y)  \, \mathcal{M}_b(\ell_2/y) \, \mathcal A_s(\ell_1/y) \, \mathcal A_s(\ell_2/y)   \nonumber  \\
&\times \left[\int d k_3 \, k_3^2 \, j_{\ell_3}(k_3 y) \, {\cal T}_{\ell_3(s)}^{X}(k_3)  \right] \, , 
\end{align}
where the baryon transfer function is evaluated at the redshift $z$.  
 
\subsection{Secondary} \label{app:computations_sec}

We start by rewriting Eq. \eqref{eq:kernel} in terms of a complete spherical harmonics angular decomposition 
\begin{align} \label{eq:kernel2}
 F_2^{(s)}(\qone,\qtwo) &= c_1 + c_2 \, (\hat q_1 \cdot \hat q_2) \left(\frac{q_1}{q_2} + \frac{q_2}{q_1} \right) + c_3 \, (\hat q_1 \cdot \hat q_2)^2 \nonumber \\
&= d_0 + \sum_{J=1}^2 \,\sqrt{\frac{4 \pi}{2 J +1}} \, Y_{J 0}(\hat q_1 \cdot \hat q_2) \,  d_J(q_1, q_2) \nonumber \\
&= d_0 + \sum_{J=1}^2 \frac{4 \pi}{2 J +1} \, \sum_{M}  \, Y_{J M}(\hat q_1)  \,  Y^*_{J M}(\hat q_2)  \,  d_J(q_1, q_2)  \, ,
\end{align}
where
\begin{align}
d_0 &= c_1 + \frac{1}{3} \, c_3 \, , \nonumber \\
d_1 &=   \,  \, c_2 \, \left(\frac{q_1}{q_2} + \frac{q_2}{q_1} \right) \,  ,\nonumber \\
d_2 &= \frac{2}{3}  \, c_3 \, \, . 
\end{align}
After this, we insert Eq. \eqref{eq:secbispectrum} into \eqref{eq:cross_full_exp}. We get
\begin{align} \label{eq:cross_full_sec} 
\langle a_{\ell_1 m_1}^{21} a_{\ell_2 m_2}^{21} a^X_{\ell_3 m_3}\rangle = & \frac{16}{\pi^3} \, i^{\ell_1 + \ell_2 + \ell_3} \, \int dr' \, \int dr'' \, W_{r(z)}(r') \, W_{r(z)}(r'') \, \int_0^\infty \, y^2 \, dy \, \left[\prod_{i=1}^3 \int d^3 k_i \, Y_{\ell_i m_i}^*(\hat{k}_i)  \right] \, \nonumber \\
& \quad \times \left[\prod_{n=1}^3 \sum_{L_n M_n} \, (-1)^{L_n/2} \,  j_{L_n}(k_i y) \,Y_{L_n M_n}^*(\hat{k}_i) \right] \, h_{L_1 L_2 L_3}^{0 0 0} \, \begin{pmatrix}
	L_1 & L_2 & L_3 \\
	M_1 & M_2 & M_3
	\end{pmatrix}  \nonumber \\
& \quad \times j_{\ell_1}(k_1 r') \, j_{\ell_2}(k_2 r'') \, {\cal T}_{\ell_3(s)}^{X}(k_3) \, \times \left(\mathcal{M}_b(k_2)\right)^2 \, \mathcal{M}_b(k_3) \, P_\zeta(k_2) \, P_\zeta(k_3) \nonumber \\
& \qquad \times  \, \Big[\alpha_1(r'') \, \Big(\alpha_2(r') d_0 +  \alpha_3(r') \Big)   \nonumber\\ 
	&\qquad  \qquad \qquad + \alpha_1(r'') \, \alpha_2(r') \sum_{j=1}^2 \frac{4 \pi}{2 J +1}\, \sum_{M}  \, Y_{J M}(\hat k_2)  \,  Y^*_{J M}(\hat k_3)  \,  d_J(k_2, k_3) \Big) \Big] \nonumber \\
  & \qquad \qquad +\ell_1 \leftrightarrow \ell_2 \, ,
\end{align} 
where we dropped the redshift dependence of the baryon transfer functions for simplicity. Again, we now separate the angular and radial integrations as 
\begin{align} \label{eq:cross_full_sec2}
\langle a_{\ell_1 m_1}^{21} a_{\ell_2 m_2}^{21} a^X_{\ell_3 m_3}\rangle = & \frac{16}{\pi^3} \, i^{\ell_1 + \ell_2 + \ell_3} \, \int dr' \, \int dr'' \, W_{r(z)}(r') \, W_{r(z)}(r'') \, \int_0^\infty \, y^2 \, dy \, \left[\prod_{i=1}^3 \int d k_i \, k_i^2 \, \right] \, \nonumber \\
& \quad \times \left[\prod_{n=1}^3 \sum_{L_n M_n} \, (-1)^{L_n/2} \,  j_{L_n}(k_i y) \right] \, h_{L_1 L_2 L_3}^{0 0 0} \, \begin{pmatrix}
	L_1 & L_2 & L_3 \\
	M_1 & M_2 & M_3
	\end{pmatrix}   \nonumber \\
& \quad \times j_{\ell_1}(k_1 r') \, j_{\ell_2}(k_2 r'') \, {\cal T}_{\ell_3(s)}^{X}(k_3) \, \times \left(\mathcal{M}_b(k_2)\right)^2 \, \mathcal{M}_b(k_3) \, P_\zeta(k_2) \, P_\zeta(k_3) \nonumber \\
& \quad \times (-1)^{m_1} \int d \hat k_1 \, Y_{\ell_1 -m_1}(\hat{k}_1) \, Y_{L_1 M_1}^*(\hat{k}_1) \times \Bigg[ \alpha_1(r'') \, \Big(\alpha_2(r') d_0 +  \alpha_3(r') \Big) \nonumber \\
& \quad \times (-1)^{m_2+m_3}  \int d \hat k_2 \, Y_{\ell_2 -m_2}(\hat{k}_2) \, Y_{L_2 M_2}^*(\hat{k}_2) \, \int d \hat k_3 \, Y_{\ell_3 -m_3}(\hat{k}_3) \, Y_{L_3 M_3}^*(\hat{k}_3)  \nonumber \\ 
& \qquad \qquad \qquad  + \alpha_1(r'') \, \alpha_2(r') \sum_{J=1}^2\, \sum_{M}  \frac{4 \pi}{2 J +1} \,  d_J(k_2, k_3) \nonumber \\
& \quad \times  \,(-1)^{M} \int d \hat k_2 \, Y_{\ell_2 m_2}^*(\hat{k}_2) \, Y_{L_2 M_2}^*(\hat{k}_2) Y_{J -M}^*(\hat k_2) \, \int d \hat k_3 \, Y_{\ell_3 m_3}^*(\hat{k}_3) \, Y_{L_3 M_3}^*(\hat{k}_3) \, Y^*_{J M}(\hat k_3) \Bigg] \nonumber \\
  & \qquad \qquad +\ell_1 \leftrightarrow \ell_2 \, .
\end{align} 
We perform the angular integrations in terms of Wigner symbols, obtaining 
\begin{align} \label{eq:cross_full_sec3}
\langle a_{\ell_1 m_1}^{21} a_{\ell_2 m_2}^{21} a^X_{\ell_3 m_3}\rangle = & \frac{16}{\pi^3} \, i^{\ell_1 + \ell_2 + \ell_3} \, \int dr' \, \int dr'' \, W_{r(z)}(r') \, W_{r(z)}(r'') \, \int_0^\infty \, y^2 \, dy \, \left[\prod_{i=1}^3 \int d k_i \, k_i^2 \, \right] \, \nonumber \\
& \quad \times \left[\prod_{n=1}^3 \sum_{L_n M_n} \, (-1)^{L_n/2} \,  j_{L_n}(k_i y) \right] \, h_{L_1 L_2 L_3}^{0 0 0} \, \begin{pmatrix}
	L_1 & L_2 & L_3 \\
	M_1 & M_2 & M_3
	\end{pmatrix}   \nonumber \\
& \quad \times j_{\ell_1}(k_1 r') \, j_{\ell_2}(k_2 r'') \, {\cal T}_{\ell_3(s)}^{X}(k_3) \, \times \left(\mathcal{M}_b(k_2)\right)^2 \, \mathcal{M}_b(k_3) \, P_\zeta(k_2) \, P_\zeta(k_3) \nonumber \\
& \quad \times (-1)^{m_1} \, \delta_{\ell_1 L_1} \, \delta_{m_1 -M_1} \times \Bigg[ \alpha_1(r'') \, \Big(\alpha_2(r') d_0 +  \alpha_3(r') \Big) \nonumber \\
& \quad \times (-1)^{m_2+m_3}  \, \delta_{\ell_2 L_2} \, \delta_{m_2 -M_2}  \, \, \delta_{\ell_3 L_3} \, \delta_{m_3 -M_3}   \nonumber \\ 
& \qquad \qquad \qquad  + \alpha_1(r'') \, \alpha_2(r') \sum_{J=1}^2\, \sum_{M}  \frac{4 \pi}{2 J +1} \,  d_J(k_2, k_3) \nonumber \\
& \quad \times  \,(-1)^{M} h_{\ell_2 L_2 J}^{0 0 0} \, h_{\ell_3 L_3 J}^{0 0 0} \, \begin{pmatrix}
	\ell_2 & L_2 & J \\
	m_2 & M_2 & -M
	\end{pmatrix} \, \begin{pmatrix}
	\ell_3 & L_3 & J \\
	m_3 & M_3 & M
	\end{pmatrix}  \Bigg] \nonumber \\
  & \qquad \qquad +\ell_1 \leftrightarrow \ell_2 \, .
\end{align} 
We can express the 3j symbols in terms of 6j symbols exploiting Eq. \eqref{eq:6j}. We get
\begin{align} \label{eq:cross_full_sec4}
\langle a_{\ell_1 m_1}^{21} a_{\ell_2 m_2}^{21} a^X_{\ell_3 m_3}\rangle = & 64 \pi \, (-1)^{\ell_1 + \ell_2 + \ell_3}  \, h_{\ell_1 \ell_2 \ell_3}^{0 0 0} \, \begin{pmatrix}
	\ell_1 & \ell_2 & \ell_3 \\
	m_1 & m_2 & m_3
	\end{pmatrix} \, \int_0^\infty \, y^2 \, dy \, \int d k_1 \, d k_2 \, d k_3 \,  \nonumber \\
& \quad \times \, j_{\ell_1}(k_1 y) \, j_{\ell_2}(k_2 y) \, j_{\ell_3}(k_3 y)  \times \, \delta_{\ell_1}(k_1) \, \beta^{\rm sec}_{\ell_2}(k_2) \, {\cal T}_{\ell_3(s)}^{X}(k_3) \, \nonumber \\
& \quad \times (\mathcal{M}_b(k_2))^2 \, \mathcal{M}_b(k_3) \times \, \frac{k_1^2}{k_2 k_3} \, \mathcal A_s(k_2) \, \mathcal A_s(k_3)  \nonumber \\
& +   64 \pi \, i^{\ell_2 + \ell_3}  \, \sqrt{\frac{(2 \ell_1+1) (2 \ell_2+1) (2 \ell_3+1)}{4 \pi}} \begin{pmatrix}
	\ell_1 & \ell_2 & \ell_3 \\
	m_1 & m_2 & m_3
	\end{pmatrix} \,  \int_0^\infty \, y^2 \, dy \, \int d k_1 \, d k_2 \, d k_3  \, \nonumber \\
& \quad \times \left[\sum_{L_2, L_3, J} \, i^{L_2+L_3} \, (-1)^J \,  \, j_{\ell_1}(k_1 y) \, j_{L_2}(k_2 y) \, j_{L_3}(k_3 y) \right] \, (2 L_2 +1) \, (2 L_3 +1)  \,    \nonumber \\
& \quad \times \epsilon_{\ell_1}(k_1) \, \beta^{\rm sec}_{\ell_2}(k_2) \, {\cal T}_{\ell_3(s)}^{X}(k_3) \, \times (\mathcal{M}_b(k_2))^2 \, \mathcal{M}_b(k_3) \times \, \frac{k_1^2}{k_2 k_3} \, \mathcal A_s(k_2) \, \mathcal A_s(k_3) \nonumber \\
& \qquad  \, \times d_J(k_2, k_3) \, \begin{pmatrix}
	\ell_1 & L_2 & L_3 \\
	0 & 0 & 0
	\end{pmatrix} \, \begin{pmatrix}
	\ell_2 & L_2 & J \\
	0 & 0 & 0
	\end{pmatrix}  \, \begin{pmatrix}
	\ell_3 & L_3 & J \\
	0 & 0 & 0
	\end{pmatrix}  \, \left\{\begin{matrix}
	\ell_1 & \ell_2 & \ell_3 \\
	J & L_3 & L_2	\end{matrix}\right\} \nonumber \\
  & \qquad \qquad +\ell_1 \leftrightarrow \ell_2 \, , 
\end{align} 
where 
\begin{align}
\beta^{\rm sec}_{\ell}(k) &= \, \int dr' \,  W_{r(z)}(r') \, \alpha_1(r') \, j_{\ell}(k r')  \, , \\
\delta_{\ell}(k) &= \, \int dr' \,  W_{r(z)}(r') \, \Big(\alpha_2(r') d_0 +  \alpha_3(r') \Big) \, j_{\ell}(k r')  \, , \\
\epsilon_{\ell}(k) &= \, \int dr' \,  W_{r(z)}(r') \, \alpha_2(r') \, j_{\ell}(k r') \, .
\end{align}
Matching Eq. \eqref{eq:cross_full_sec4} with \eqref{eq:decisot}, we read our angular averaged secondary contribution
\begin{align} \label{eq:cross_full_sec5}
B^{21-21-\rm X, \, sec}_{\ell_1 \ell_2 \ell_3}(z) = & 64 \pi \, (-1)^{\ell_1 + \ell_2 + \ell_3}  \, h_{\ell_1 \ell_2 \ell_3}^{0 0 0} \, \int_0^\infty \, y^2 \, dy \, \int d k_1 \, d k_2 \, d k_3 \,  \nonumber \\
& \quad \times \, j_{\ell_1}(k_1 y) \, j_{\ell_2}(k_2 y) \, j_{\ell_3}(k_3 y)  \times \, \delta_{\ell_1}(k_1) \, \beta^{\rm sec}_{\ell_2}(k_2) \, {\cal T}_{\ell_3(s)}^{X}(k_3) \, \nonumber \\
& \quad \times (\mathcal{M}_b(k_2))^2 \, \mathcal{M}_b(k_3) \times \, \frac{k_1^2}{k_2 k_3} \, \mathcal A_s(k_2) \, \mathcal A_s(k_3)  \nonumber \\
& +   64 \pi \, i^{\ell_2 + \ell_3}  \, \sqrt{\frac{(2 \ell_1+1) (2 \ell_2+1) (2 \ell_3+1)}{4 \pi}}  \,  \int_0^\infty \, y^2 \, dy \, \int d k_1 \, d k_2 \, d k_3  \, \nonumber \\
& \quad \times \left[\sum_{L_2, L_3, J} \, i^{L_2+L_3} \, (-1)^J \,  \, j_{\ell_1}(k_1 y) \, j_{L_2}(k_2 y) \, j_{L_3}(k_3 y) \right] \, (2 L_2 +1) \, (2 L_3 +1)  \,    \nonumber \\
& \quad \times \epsilon_{\ell_1}(k_1) \, \beta^{\rm sec}_{\ell_2}(k_2) \, {\cal T}_{\ell_3(s)}^{X}(k_3) \, \times (\mathcal{M}_b(k_2))^2 \, \mathcal{M}_b(k_3) \times \, \frac{k_1^2}{k_2 k_3} \, \mathcal A_s(k_2) \, \mathcal A_s(k_3) \nonumber \\
& \qquad  \, \times d_J(k_2, k_3) \, \begin{pmatrix}
	\ell_1 & L_2 & L_3 \\
	0 & 0 & 0
	\end{pmatrix} \, \begin{pmatrix}
	\ell_2 & L_2 & J \\
	0 & 0 & 0
	\end{pmatrix}  \, \begin{pmatrix}
	\ell_3 & L_3 & J \\
	0 & 0 & 0
	\end{pmatrix}  \, \left\{\begin{matrix}
	\ell_1 & \ell_2 & \ell_3 \\
	J & L_3 & L_2	\end{matrix}\right\} \nonumber \\
  & \qquad \qquad +\ell_1 \leftrightarrow \ell_2 \, . 
\end{align} 
Again, we can reorder the final result factorizing the momenta integration as
\begin{align} \label{eq:cross_full_sec6}
& B^{21-21-\rm X, \, sec}_{\ell_1 \ell_2 \ell_3}(z) = 64 \pi \,  \sqrt{\frac{(2 \ell_1+1) (2 \ell_2+1) (2 \ell_3+1)}{4 \pi}} \,    \nonumber \\
 & \times  \Bigg[(-1)^{\ell_1 + \ell_2 + \ell_3} \, \begin{pmatrix}
	\ell_1 & \ell_2 & \ell_3 \\
	0 & 0 & 0
	\end{pmatrix} \,  \mathcal I^{\rm sec, 1}_{\ell_1 \ell_2 \ell_3}   \nonumber \\
& \qquad \qquad - c_2  \, i^{\ell_2 + \ell_3}  \, \sum_{L_2 L_3} \, i^{L_2 + L_3} \, (2 L_2 +1) \, (2 L_3 +1) \, \, \begin{pmatrix}
	\ell_1 & L_2 & L_3 \\
	0 & 0 & 0
	\end{pmatrix} \, \begin{pmatrix}
	\ell_2 & L_2 & 1 \\
	0 & 0 & 0
	\end{pmatrix}  \, \begin{pmatrix}
	\ell_3 & L_3 & 1 \\
	0 & 0 & 0
	\end{pmatrix}  \, \left\{\begin{matrix}
	\ell_1 & \ell_2 & \ell_3 \\
	1 & L_3 & L_2	\end{matrix}\right\} \,   \nonumber \\
& \qquad \qquad \qquad \qquad \times \left(\mathcal I^{\rm sec, 2}_{\ell_1 \ell_2 \ell_3, \, L_2 L_3} + \mathcal I^{\rm sec, 3}_{\ell_1 \ell_2 \ell_3, \, L_2 L_3}\right) \nonumber \\
& \qquad \qquad + \frac{2}{3}  c_3 \, i^{\ell_2 + \ell_3} \sum_{L_2 L_3} \, i^{L_2 + L_3} \, (2 L_2 +1) \, (2 L_3 +1) \, \, \begin{pmatrix}
	\ell_1 & L_2 & L_3 \\
	0 & 0 & 0
	\end{pmatrix} \, \begin{pmatrix}
	\ell_2 & L_2 & 2 \\
	0 & 0 & 0
	\end{pmatrix}  \, \begin{pmatrix}
	\ell_3 & L_3 & 2 \\
	0 & 0 & 0
	\end{pmatrix}  \, \left\{\begin{matrix}
	\ell_1 & \ell_2 & \ell_3 \\
	2 & L_3 & L_2	\end{matrix}\right\} \nonumber  \\
&  \qquad \qquad \qquad \qquad  \times \left(\mathcal I^{\rm sec, 4}_{\ell_1 \ell_2 \ell_3, \, L_2 L_3}\right) \, \Bigg] \nonumber \\
& \qquad \qquad \qquad \qquad \qquad \qquad \qquad \qquad \qquad \qquad  + \ell_1 \leftrightarrow \ell_2 \, ,
\end{align}
where
\begin{align} \label{eq:Isec_1}
\mathcal I^{\rm sec, 1}_{\ell_1 \ell_2 \ell_3} = \,\int_0^\infty \, & y^2 \, dy \left[\int d k_1 \, k^2_1 \, j_{\ell_1}(k_1 y) \, \delta_{\ell_1}(k_1)   \right] \nonumber  \\
&\times \left[\int \frac{d k_2}{k_2}\, j_{\ell_2}(k_2 y) \, \beta^{\rm sec}_{\ell_2}(k_2) \, (\mathcal{M}_b(k_{2}))^2 \,\mathcal A_s(k_2) \right] \nonumber \\
&\times \left[\int \frac{d k_3}{k_3} \, j_{\ell_3}(k_3 y) \, {\cal T}_{\ell_3(s)}^{X}(k_3) \, \mathcal{M}_b(k_{3})  \,\mathcal A_s(k_3) \right] \, , 
\end{align}
\begin{align}
&\mathcal I^{\rm sec, 2}_{\ell_1 \ell_2 \ell_3, \, L_2 L_3} =  \,  \int_0^\infty \, y^2 \, dy \left[\int d k_1 \, k^2_1 \, j_{\ell_1}(k_1 y) \, \epsilon_{\ell_1}(k_1)   \right] \nonumber  \\
& \qquad \qquad \qquad \qquad \times \left[\int d k_2 \, j_{L_2}(k_2 y) \, \beta^{\rm sec}_{\ell_2}(k_2) \, (\mathcal{M}_b(k_{2}))^2 \,\mathcal A_s(k_2) \right] \nonumber \\
&\qquad \qquad \qquad \qquad  \times \left[\int \frac{d k_3}{k_3^2} \, j_{L_3}(k_3 y) \, {\cal T}_{\ell_3(s)}^{X}(k_3) \, \mathcal{M}_b(k_{3})  \,\mathcal A_s(k_3) \right] \, , 
\end{align}
\begin{align}
&\mathcal I^{\rm sec, 3}_{\ell_1 \ell_2 \ell_3, \, L_2 L_3} = \int_0^\infty \, y^2 \, dy \left[\int d k_1 \, k^2_1 \, j_{\ell_1}(k_1 y) \, \epsilon_{\ell_1}(k_1)   \right] \nonumber  \\
& \qquad \qquad \qquad \qquad \times \left[\int \frac{d k_2}{k_2^2} \, j_{L_2}(k_2 y) \, \beta^{\rm sec}_{\ell_2}(k_2) \, (\mathcal{M}_b(k_{2}))^2 \,\mathcal A_s(k_2) \right] \nonumber \\
&\qquad \qquad \qquad \qquad  \times \left[\int d k_3 \, j_{L_3}(k_3 y) \, {\cal T}_{\ell_3(s)}^{X}(k_3) \, \mathcal{M}_b(k_{3})  \,\mathcal A_s(k_3) \right] \, , 
\end{align}
\begin{align} \label{eq:Isec_4}
&\mathcal I^{\rm sec, 4}_{\ell_1 \ell_2 \ell_3, \, L_2 L_3} = \int_0^\infty \, y^2 \, dy \left[\int d k_1 \, k^2_1 \, j_{\ell_1}(k_1 y) \, \epsilon_{\ell_1}(k_1)   \right] \nonumber  \\
&  \qquad \qquad \qquad \qquad \times \left[\int \frac{d k_2}{k_2} \, j_{L_2}(k_2 y) \, \beta^{\rm sec}_{\ell_2}(k_2) \, (\mathcal{M}_b(k_{2}))^2 \,\mathcal A_s(k_2) \right] \nonumber \\
&\qquad \qquad \qquad \qquad  \times \left[\int \frac{d k_3}{k_3} \, j_{L_3}(k_3 y) \, {\cal T}_{\ell_3(s)}^{X}(k_3) \, \mathcal{M}_b(k_{3})  \,\mathcal A_s(k_3) \right] \, . 
\end{align}

\subsubsection*{Limber approximation}

We can derive the following approximated formulas for the integrals \eqref{eq:Isec_1}-\eqref{eq:Isec_4} by applying the Limber approximation 
\begin{align} \label{eq:I1limberm}
\mathcal I^{\rm sec, 1}_{\ell_1 \ell_2 \ell_3} = \left(\frac{\pi}{2}\right)^2 \, \left(\frac{1}{\ell_2}\right)^3 \, \alpha_1(z) \, \Big[\alpha_2(z) d_0 +  \alpha_3(z) \Big] \, \int_0^\infty  \,  dy \, & y \, \left[W_{r(z)}(y)\right]^2 \,  \, \left[\mathcal{M}_b(\ell_2/y)\right]^2 \, \mathcal A_s(\ell_2/y)  \nonumber \\
&\times \left[\int \frac{d k_3}{k_3} \, j_{\ell_3}(k_3 y) \, {\cal T}_{\ell_3(s)}^{X}(k_3) \, \mathcal{M}_b(k_{3})  \,\mathcal A_s(k_3) \right] \, , 
\end{align}
\begin{align}
&\mathcal I^{\rm sec, 2}_{\ell_1 \ell_2 \ell_3, \, L_2 L_3} =  \left(\frac{\pi}{2}\right)^2 \, \left(\frac{1}{\ell_2 \, (L_2)^2}\right) \,  \alpha_1(z) \, \alpha_2(z) \, \int_0^\infty  \,  dy  \, \left[W_{r(z)}(y)\right] \, \left[W_{r(z)}\left(\frac{\ell_2}{L_2} \, y\right)\right] \, \, \nonumber \\
&\qquad \qquad \qquad \qquad \qquad \qquad\qquad \qquad\qquad \qquad \qquad \qquad  \times \, \left[\mathcal{M}_b(L_2/y)\right]^2 \, \mathcal A_s(L_2/y) \nonumber \\
&\qquad \qquad \qquad \qquad \qquad \qquad \qquad\qquad \qquad \qquad \qquad  \times \left[\int \frac{d k_3}{k_3^2} \, j_{L_3}(k_3 y) \, {\cal T}_{\ell_3(s)}^{X}(k_3) \, \mathcal{M}_b(k_{3})  \,\mathcal A_s(k_3) \right] \, , 
\end{align}
\begin{align}
&\mathcal I^{\rm sec, 3}_{\ell_1 \ell_2 \ell_3, \, L_2 L_3} = \left(\frac{\pi}{2}\right)^2 \, \left(\frac{1}{\ell_2 \, (L_2)^4}\right) \,  \alpha_1(z) \, \alpha_2(z) \, \int_0^\infty  \,  dy  \, y^2 \, \left[W_{r(z)}(y)\right] \, \left[W_{r(z)}\left(\frac{\ell_2}{L_2} \, y\right)\right] \,  \,   \nonumber \\
&\qquad \qquad \qquad \qquad\qquad \qquad \qquad \qquad \qquad \qquad \qquad \qquad  \times \, \left[\mathcal{M}_b(L_2/y)\right]^2 \, \mathcal A_s(L_2/y) \nonumber \\
&\qquad \qquad \qquad \qquad \qquad\qquad \qquad\qquad \qquad \qquad \qquad  \times \left[\int d k_3 \, j_{L_3}(k_3 y) \, {\cal T}_{\ell_3(s)}^{X}(k_3) \, \mathcal{M}_b(k_{3})  \,\mathcal A_s(k_3) \right] \, , 
\end{align}
\begin{align}  \label{eq:I4limberm}
&\mathcal I^{\rm sec, 4}_{\ell_1 \ell_2 \ell_3, \, L_2 L_3} = \left(\frac{\pi}{2}\right)^2 \, \left(\frac{1}{\ell_2 \, (L_2)^3}\right) \,  \alpha_1(z) \, \alpha_2(z) \, \int_0^\infty  \,  dy  \, y \, \left[W_{r(z)}(y)\right] \, \left[W_{r(z)}\left(\frac{\ell_2}{L_2} \, y\right)\right]  \,   \nonumber \\
&\qquad \qquad \qquad \qquad\qquad \qquad \qquad \qquad \qquad \qquad \qquad \qquad  \times \, \left[\mathcal{M}_b(L_2/y)\right]^2 \, \mathcal A_s(L_2/y) \nonumber \\
&\qquad \qquad \qquad\qquad \qquad \qquad \qquad \qquad \qquad \qquad \qquad \times \left[\int \frac{d k_3}{k_3} \, j_{L_3}(k_3 y) \, {\cal T}_{\ell_3(s)}^{X}(k_3) \, \mathcal{M}_b(k_{3})  \,\mathcal A_s(k_3) \right] \, . 
\end{align}

\section{Spin-weighted spherical harmonics} \label{app:Wigner}

In this appendix, we give some useful formulas for spin-weighted spherical harmonics and their integration. We will use $(\theta, \phi)$ or $\hat x$ to denote a given direction on the $2D$ sphere and $d^2 \hat n$ or $d^2 \Omega_x$ to indicate the infinitesimal solid angle on the sphere. We will also review some technical computations of this work. The formulas we provide here allow us to simplify the expressions for spherical harmonic coefficients when dealing with primordial perturbations from inflation. We refer the reader to e.g. \cite{Okamoto:2002ik,Komatsu:2003iq, Liguori:2005rj,Shiraishi:2012bh} for more details.

\subsection*{Basics}

We start with the orthogonality and completeness conditions for the spin-weighted spherical harmonics ${}_sY_{\ell m}(\hat x)$ 
\begin{align} \label{eq:orto_harm}
\int d^2 \Omega_x \,\, {}_s Y_{\ell m}^*(\hat x)
\, {}_s Y_{\ell' m'}(\hat x) &= \delta_{\ell, \ell'} \, \delta_{m, m'} \, , \nonumber \\
\sum_{\ell m} \, \, {}_s Y_{\ell m}^*(\hat x) \,
{}_s Y_{\ell m}(\hat x')
&= \delta(\hat x - \hat x') \, ,
\end{align} 
as well as the following properties regarding the transformations under conjugation and parity
\begin{align}
{}_sY^*_{\ell m}(\theta, \phi) &= (-1)^{s + m}{}_{-s}Y_{\ell -m}(\theta, \phi) \, , \nonumber \\
{}_s Y_{\ell m}(\pi - \theta, \phi + \pi) &= (-1)^{\ell} \, {}_{-s} Y_{\ell m}(\theta, \phi) \, . 
\label{eq:parity_Y}\end{align}
We can decompose the weighted spherical harmonics evaluated at an angle between two vectors $\hat k \cdot \hat q$ as (see e.g. \cite{Okamoto:2002ik})
\begin{align} \label{eq:gen_add_rel}
{}_{s} Y_{\ell m}(\hat k \cdot \hat q) = &\sqrt{\frac{4 \pi}{2 \ell +1}} \, (-1)^s \, \sum_{M}  \, {}_{s} Y_{\ell M}(\hat k)  \,  {}_{-m} Y^*_{\ell M}(\hat q)  \nonumber \\
 = &\sqrt{\frac{4 \pi}{2 \ell +1}} \, (-1)^s \, \sum_{M}  \, {}_{s} Y_{\ell M}(\hat q)  \,  {}_{-m} Y^*_{\ell M}(\hat k) \, ,
\end{align}
which is a variation of the so-called generalized addition relation. 

Another important result is the plane-wave decomposition in terms of spin-0 spherical harmonics 
\begin{align} \label{eq:plane-wave}
e^{i \vec q \cdot \vec x} = & \sum_\ell \sqrt{4 \pi (2 \ell +1)} \, i^\ell  \, j_\ell(q x) \, Y_{\ell 0}(\hat x \cdot \hat q) \nonumber \\
= & \sum_\ell 4 \pi \, i^\ell  \, j_\ell(q x)  \sum_M \, Y_{\ell M}(\hat q) \, Y^*_{\ell M}(\hat x)\, .
\end{align}
As a last useful equation, we give the Clebsch-Gordan relation
\begin{align} \label{eq:rel_wigner}
\prod_{i = 1}^2 {}_{s_i}Y_{\ell_i m_i}(\hat x)&= \sum_{\ell_3 m_3 s_3} \,  {}_{s_3}Y^*_{\ell_3 m_3}(\hat x) \, \sqrt{\frac{(2 \ell_1 + 1) (2 \ell_2 + 1) (2 \ell_3 + 1)}{4 \pi}}   \nonumber\\
& \times \begin{pmatrix}
	\ell_1 & \ell_2 & \ell_3 \\
	-s_1 & -s_2 & -s_3
	\end{pmatrix}\begin{pmatrix}
	\ell_1 & \ell_2 & \ell_3 \\
	m_1 & m_2 & m_3
	\end{pmatrix} \, ,
\end{align}
which can be used to compose the angular momenta of two separate spherical harmonics evaluated at the same angle. Together with \eqref{eq:plane-wave}, we can employ this result to isolate the radial and angular dependencies of a given expression (see e.g. \cite{Hu:1997hp} for more on this aspect). 

In Eq. \eqref{eq:rel_wigner} we have introduced the Wigner 3-j symbols, which are related to the well-known Clebsch-Gordan coefficients
\begin{equation} \label{eq:CG}
\mathcal C^{\ell_2 m_3}_{\ell_1 m_1 \ell_2 m_2} = \langle \ell_1  m_1 \ell_2 m_2| \ell_3 m_3  \rangle
\end{equation}
through (see e.g. \cite{Shiraishi:2012bh})
\begin{equation} \label{eq:3jsymbols-CG}
\begin{pmatrix}
	\ell_1 & \ell_2 & \ell_3 \\
	m_1 & m_2 & - m_3
	\end{pmatrix} =  \frac{(-1)^{\ell_1 - \ell_2 + m_3}}{\sqrt{2 \ell_3 + 1}} \, \mathcal C^{\ell_2 m_3}_{\ell_1 m_1 \ell_2 m_2} \, .
\end{equation}
Therefore, the 3-j symbols of the form \eqref{eq:3jsymbols-CG} vanish unless the selection rules are satisfied as follows
\begin{align}
&|m_1| \leq \ell_1 \,, \qquad  |m_2|\leq \ell_2 \,, \qquad |m_3|\leq \ell_3 \,, \qquad  m_1 +m_2 =m_3 \, , \nonumber \\
&|\ell_1 − \ell_2| \leq \ell_3 \leq \ell_1 + \ell_2 \quad \mbox{(the triangle condition)} \, , \qquad \ell_1 +\ell_2 + \ell_3 \in Z \, .
\end{align}
Some useful properties of the Wigner 3-j symbols are the following transformation rules under the $m_i$-sign inversion and odd permutations of columns 
\begin{align} \label{eq:lwigner}
 \begin{pmatrix}
	\ell_1 & \ell_2 & \ell_3 \\
	m_1 & m_2 & m_3
	\end{pmatrix} =& (-1)^{\sum_i \ell_i}\begin{pmatrix}
	\ell_1 & \ell_2 & \ell_3 \\
	 - m_1 & - m_2 & - m_3
	\end{pmatrix}  \nonumber \\
	=& (-1)^{\sum_i \ell_i}\begin{pmatrix}
	\ell_2 & \ell_1 & \ell_3 \\
	  m_2 &  m_1 & m_3
	\end{pmatrix} \, .
\end{align}
These symbols are left invariant by even permutations of columns. 

Another useful property of the Wigner 3-j symbols is the orthogonality condition
\begin{align} \label{eq:sum_m_wigner}
\sum_{m_1, m_2}  \begin{pmatrix}
	\ell_1 & \ell_2 & \ell_3 \\
	m_1 & m_2 & m_3
	\end{pmatrix}\begin{pmatrix}
	\ell_1 & \ell_2 & \ell'_3 \\
	m_1 & m_2 & m'_3
	\end{pmatrix}  = (2 \ell_3 + 1)^{-1} \, \delta_{\ell_3, \ell'_3} \, \delta_{m_3, m'_3} \, .
\end{align}
We can express the product of 3j symbols in terms of 6j-symbols as
\begin{align} \label{eq:6j}
\sum_{m_4, m_5, m_6} & (-1)^{\ell_4+\ell_5+\ell_6-m_4-m_5-m_6} \begin{pmatrix}
	\ell_5 & \ell_1 & \ell_6 \\
	m_5 & - m_1 & - m_6
	\end{pmatrix} 
    \begin{pmatrix}
	\ell_6 & \ell_2 & \ell_4 \\
	m_6 & -m_2 & -m_4
	\end{pmatrix} 
    \begin{pmatrix}
	\ell_4 & \ell_3 & \ell_5 \\
	m_4 & -m_3 & -m_5
	\end{pmatrix} \nonumber \\
& \qquad = \begin{pmatrix}
	\ell_1 & \ell_2 & \ell_3 \\
	m_1 & m_2 & m_3
	\end{pmatrix} \, \left\{\begin{matrix}
	\ell_1 & \ell_2 & \ell_3 \\
	\ell_4 & \ell_5 & \ell_6
	\end{matrix}\right\}  \,  .
\end{align}
More properties of the Wigner symbols can be found in \cite{Shiraishi:2012bh}.

\subsection*{Integration}

We define the quantity ${}_{s_1 s_2 s_3} \mathcal G_{\ell_1 \ell_2 \ell_3}^{m_1 m_2 m_3}$, which is known as ``generalized" Gaunt integral and it represents the angular integral of the product of three (weighted) spherical harmonics. This can be written in terms of Wigner 3-j symbols as (see e.g. \cite{Komatsu:2003iq, Liguori:2005rj})
\begin{align} \label{eq:Gaunt_integral}
{}_{s_1 s_2 s_3}\mathcal G_{\ell_1 \ell_2 \ell_3}^{m_1 m_2 m_3} &= \int d^2 \Omega_x \, {}_{s_1}Y_{\ell_1 m_1}(\hat x) \, {}_{s_2}Y_{\ell_2 m_2}(\hat x)  \,{}_{s_3}Y_{\ell_3 m_3}(\hat x) \nonumber\\
&= \sqrt{\frac{(2 \ell_1 + 1) (2 \ell_2 + 1) (2 \ell_3 + 1)}{4 \pi}}   \begin{pmatrix}
	\ell_1 & \ell_2 & \ell_3 \\
	-s_1 & -s_2 & -s_3
	\end{pmatrix}\begin{pmatrix}
	\ell_1 & \ell_2 & \ell_3 \\
	m_1 & m_2 & m_3
	\end{pmatrix} \, .
\end{align}

\FloatBarrier
\bibliographystyle{utcaps}
\bibliography{References}

\providecommand{\href}[2]{#2}\begingroup\raggedright\begin{thebibliography}{10}

\bibitem{Planck:2018vyg}
The {\bfseries Planck}, N.~Aghanim {\em et~al.}, ``{Planck 2018 results. VI.
  Cosmological parameters}'',
  \href{http://dx.doi.org/10.1051/0004-6361/201833910}{{\em Astron. Astrophys.}
  {\bfseries 641} (2020) A6}, \href{http://arxiv.org/abs/1807.06209}{{\ttfamily
  arXiv:1807.06209 [astro-ph.CO]}}. [Erratum: Astron.Astrophys. 652, C4
  (2021)].

\bibitem{Planck:2019kim}
The {\bfseries Planck}, Y.~Akrami {\em et~al.}, ``{Planck 2018 results. IX.
  Constraints on primordial non-Gaussianity}'',
  \href{http://dx.doi.org/10.1051/0004-6361/201935891}{{\em Astron. Astrophys.}
  {\bfseries 641} (2020) A9}, \href{http://arxiv.org/abs/1905.05697}{{\ttfamily
  arXiv:1905.05697 [astro-ph.CO]}}.

\bibitem{Maldacena:2002vr}
J.~M. Maldacena, ``{Non-Gaussian features of primordial fluctuations in single
  field inflationary models}'',
  \href{http://dx.doi.org/10.1088/1126-6708/2003/05/013}{{\em JHEP} {\bfseries
  05} (2003) 013}, \href{http://arxiv.org/abs/astro-ph/0210603}{{\ttfamily
  arXiv:astro-ph/0210603}}.

\bibitem{Cooray:2006km}
A.~Cooray, ``{21-cm Background Anisotropies Can Discern Primordial
  Non-Gaussianity}'',
  \href{http://dx.doi.org/10.1103/PhysRevLett.97.261301}{{\em Phys. Rev. Lett.}
  {\bfseries 97} (2006) 261301},
  \href{http://arxiv.org/abs/astro-ph/0610257}{{\ttfamily
  arXiv:astro-ph/0610257}}.

\bibitem{Pillepich:2006fj}
A.~Pillepich, C.~Porciani, and S.~Matarrese, ``{The bispectrum of redshifted
  21-cm fluctuations from the dark ages}'',
  \href{http://dx.doi.org/10.1086/517963}{{\em Astrophys. J.} {\bfseries 662}
  (2007) 1--14}, \href{http://arxiv.org/abs/astro-ph/0611126}{{\ttfamily
  arXiv:astro-ph/0611126}}.

\bibitem{Meerburg:2016zdz}
P.~D. Meerburg, M.~M\"unchmeyer, J.~B. Mu\~noz, and X.~Chen, ``{Prospects for
  Cosmological Collider Physics}'',
  \href{http://dx.doi.org/10.1088/1475-7516/2017/03/050}{{\em JCAP} {\bfseries
  03} (2017) 050}, \href{http://arxiv.org/abs/1610.06559}{{\ttfamily
  arXiv:1610.06559 [astro-ph.CO]}}.

\bibitem{Munoz:2015eqa}
J.~B. Mu\~noz, Y.~Ali-Ha\"\i{}moud, and M.~Kamionkowski, ``{Primordial
  non-gaussianity from the bispectrum of 21-cm fluctuations in the dark
  ages}'', \href{http://dx.doi.org/10.1103/PhysRevD.92.083508}{{\em Phys. Rev.
  D} {\bfseries 92} no.~8, (2015) 083508},
  \href{http://arxiv.org/abs/1506.04152}{{\ttfamily arXiv:1506.04152
  [astro-ph.CO]}}.

\bibitem{Silk:2020bsr}
J.~Silk, ``{The limits of cosmology: role of the Moon}'',
  \href{http://dx.doi.org/10.1098/rsta.2019.0561}{{\em Phil. Trans. A. Math.
  Phys. Eng. Sci.} {\bfseries 379} (2021) 20190561},
  \href{http://arxiv.org/abs/2011.04671}{{\ttfamily arXiv:2011.04671
  [astro-ph.CO]}}.

\bibitem{Floss:2022grj}
T.~Fl\"oss, T.~de~Wild, P.~D. Meerburg, and L.~V.~E. Koopmans, ``{The Dark
  Ages' 21-cm trispectrum}'',
  \href{http://dx.doi.org/10.1088/1475-7516/2022/06/020}{{\em JCAP} {\bfseries
  06} no.~06, (2022) 020}, \href{http://arxiv.org/abs/2201.08843}{{\ttfamily
  arXiv:2201.08843 [astro-ph.CO]}}.

\bibitem{Cole:2019zhu}
P.~S. Cole and J.~Silk, ``{Small-scale primordial fluctuations in the 21 cm
  Dark Ages signal}'', \href{http://dx.doi.org/10.1093/mnras/staa3638}{{\em
  Mon. Not. Roy. Astron. Soc.} {\bfseries 501} no.~2, (2021) 2627--2634},
  \href{http://arxiv.org/abs/1912.02171}{{\ttfamily arXiv:1912.02171
  [astro-ph.CO]}}.

\bibitem{Biagetti:2021tua}
M.~Biagetti, L.~Castiblanco, J.~Nore\~na, and E.~Sefusatti, ``{The covariance
  of squeezed bispectrum configurations}'',
  \href{http://dx.doi.org/10.1088/1475-7516/2022/09/009}{{\em JCAP} {\bfseries
  09} (2022) 009}, \href{http://arxiv.org/abs/2111.05887}{{\ttfamily
  arXiv:2111.05887 [astro-ph.CO]}}.

\bibitem{Floss:2022wkq}
T.~Fl\"oss, M.~Biagetti, and P.~D. Meerburg, ``{Primordial non-Gaussianity and
  non-Gaussian covariance}'',
  \href{http://dx.doi.org/10.1103/PhysRevD.107.023528}{{\em Phys. Rev. D}
  {\bfseries 107} no.~2, (2023) 023528},
  \href{http://arxiv.org/abs/2206.10458}{{\ttfamily arXiv:2206.10458
  [astro-ph.CO]}}.

\bibitem{Baumann:2018muz}
D.~Baumann, ``{Primordial Cosmology}'',
  \href{http://dx.doi.org/10.22323/1.305.0009}{{\em PoS} {\bfseries TASI2017}
  (2018) 009}, \href{http://arxiv.org/abs/1807.03098}{{\ttfamily
  arXiv:1807.03098 [hep-th]}}.

\bibitem{Byrnes:2010em}
C.~T. Byrnes and K.-Y. Choi, ``{Review of local non-Gaussianity from
  multi-field inflation}'', \href{http://dx.doi.org/10.1155/2010/724525}{{\em
  Adv. Astron.} {\bfseries 2010} (2010) 724525},
  \href{http://arxiv.org/abs/1002.3110}{{\ttfamily arXiv:1002.3110
  [astro-ph.CO]}}.

\bibitem{Shiraishi:2010sm}
M.~Shiraishi, S.~Yokoyama, K.~Ichiki, and K.~Takahashi, ``{Analytic formulae of
  the CMB bispectra generated from non-Gaussianity in the tensor and vector
  perturbations}'', \href{http://dx.doi.org/10.1103/PhysRevD.82.103505}{{\em
  Phys. Rev. D} {\bfseries 82} (2010) 103505},
  \href{http://arxiv.org/abs/1003.2096}{{\ttfamily arXiv:1003.2096
  [astro-ph.CO]}}.

\bibitem{Shiraishi:2010kd}
M.~Shiraishi, D.~Nitta, S.~Yokoyama, K.~Ichiki, and K.~Takahashi, ``{CMB
  Bispectrum from Primordial Scalar, Vector and Tensor non-Gaussianities}'',
  \href{http://dx.doi.org/10.1143/PTP.125.795}{{\em Prog. Theor. Phys.}
  {\bfseries 125} (2011) 795--813},
  \href{http://arxiv.org/abs/1012.1079}{{\ttfamily arXiv:1012.1079
  [astro-ph.CO]}}.

\bibitem{camb_notes}
A.~Lewis, ``{CAMB Notes}.'' \url{https://cosmologist.info/notes/CAMB.pdf}.

\bibitem{Meerburg:2013dua}
P.~D. Meerburg, C.~Dvorkin, and D.~N. Spergel, ``{Probing Patchy Reionization
  through \ensuremath{\tau}-21 cm Correlation Statistics}'',
  \href{http://dx.doi.org/10.1088/0004-637X/779/2/124}{{\em Astrophys. J.}
  {\bfseries 779} (2013) 124}, \href{http://arxiv.org/abs/1303.3887}{{\ttfamily
  arXiv:1303.3887 [astro-ph.CO]}}.

\bibitem{Lewis:2007kz}
A.~Lewis and A.~Challinor, ``{The 21cm angular-power spectrum from the dark
  ages}'', \href{http://dx.doi.org/10.1103/PhysRevD.76.083005}{{\em Phys. Rev.
  D} {\bfseries 76} (2007) 083005},
  \href{http://arxiv.org/abs/astro-ph/0702600}{{\ttfamily
  arXiv:astro-ph/0702600}}.

\bibitem{Bernardeau:2010ac}
F.~Bernardeau, C.~Pitrou, and J.-P. Uzan, ``{CMB spectra and bispectra
  calculations: making the flat-sky approximation rigorous}'',
  \href{http://dx.doi.org/10.1088/1475-7516/2011/02/015}{{\em JCAP} {\bfseries
  02} (2011) 015}, \href{http://arxiv.org/abs/1012.2652}{{\ttfamily
  arXiv:1012.2652 [astro-ph.CO]}}.

\bibitem{Hanson:2009kg}
D.~Hanson, K.~M. Smith, A.~Challinor, and M.~Liguori, ``{CMB lensing and
  primordial non-Gaussianity}'',
  \href{http://dx.doi.org/10.1103/PhysRevD.80.083004}{{\em Phys. Rev. D}
  {\bfseries 80} (2009) 083004},
  \href{http://arxiv.org/abs/0905.4732}{{\ttfamily arXiv:0905.4732
  [astro-ph.CO]}}.

\bibitem{Babich:2004yc}
D.~Babich and M.~Zaldarriaga, ``{Primordial bispectrum information from CMB
  polarization}'', \href{http://dx.doi.org/10.1103/PhysRevD.70.083005}{{\em
  Phys. Rev. D} {\bfseries 70} (2004) 083005},
  \href{http://arxiv.org/abs/astro-ph/0408455}{{\ttfamily
  arXiv:astro-ph/0408455}}.

\bibitem{Smith:2011rm}
T.~L. Smith, M.~Kamionkowski, and B.~D. Wandelt, ``{The Probability
  Distribution for Non-Gaussianity Estimators}'',
  \href{http://dx.doi.org/10.1103/PhysRevD.84.063013}{{\em Phys. Rev. D}
  {\bfseries 84} (2011) 063013},
  \href{http://arxiv.org/abs/1104.0930}{{\ttfamily arXiv:1104.0930
  [astro-ph.CO]}}.

\bibitem{Smith:2012ta}
T.~L. Smith, D.~Grin, and M.~Kamionkowski, ``{Improved estimator for
  non-Gaussianity in cosmic microwave background observations}'',
  \href{http://dx.doi.org/10.1103/PhysRevD.87.063003}{{\em Phys. Rev. D}
  {\bfseries 87} (2013) 063003},
  \href{http://arxiv.org/abs/1211.3417}{{\ttfamily arXiv:1211.3417
  [astro-ph.CO]}}.

\bibitem{Lewis:2011fk}
A.~Lewis, A.~Challinor, and D.~Hanson, ``{The shape of the CMB lensing
  bispectrum}'', \href{http://dx.doi.org/10.1088/1475-7516/2011/03/018}{{\em
  JCAP} {\bfseries 03} (2011) 018},
  \href{http://arxiv.org/abs/1101.2234}{{\ttfamily arXiv:1101.2234
  [astro-ph.CO]}}.

\bibitem{Burns:2021ndk}
J.~Burns {\em et~al.}, ``{Global 21-cm Cosmology from the Farside of the
  Moon}'', \href{http://arxiv.org/abs/2103.05085}{{\ttfamily arXiv:2103.05085
  [astro-ph.CO]}}.

\bibitem{Pober:2013jna}
J.~C. Pober {\em et~al.}, ``{What Next-Generation 21 cm Power Spectrum
  Measurements Can Teach Us About the Epoch of Reionization}'',
  \href{http://dx.doi.org/10.1088/0004-637X/782/2/66}{{\em Astrophys. J.}
  {\bfseries 782} (2014) 66}, \href{http://arxiv.org/abs/1310.7031}{{\ttfamily
  arXiv:1310.7031 [astro-ph.CO]}}.

\bibitem{Pober:2014lva}
J.~C. Pober, ``{The Impact of Foregrounds on Redshift Space Distortion
  Measurements With the Highly-Redshifted 21 cm Line}'',
  \href{http://dx.doi.org/10.1093/mnras/stu2575}{{\em Mon. Not. Roy. Astron.
  Soc.} {\bfseries 447} no.~2, (2015) 1705--1712},
  \href{http://arxiv.org/abs/1411.2050}{{\ttfamily arXiv:1411.2050
  [astro-ph.CO]}}.

\bibitem{Okamoto:2002ik}
T.~Okamoto and W.~Hu, ``{The angular trispectra of CMB temperature and
  polarization}'', \href{http://dx.doi.org/10.1103/PhysRevD.66.063008}{{\em
  Phys. Rev. D} {\bfseries 66} (2002) 063008},
  \href{http://arxiv.org/abs/astro-ph/0206155}{{\ttfamily
  arXiv:astro-ph/0206155}}.

\bibitem{Komatsu:2003iq}
E.~Komatsu, D.~N. Spergel, and B.~D. Wandelt, ``{Measuring primordial
  non-Gaussianity in the cosmic microwave background}'',
  \href{http://dx.doi.org/10.1086/491724}{{\em Astrophys. J.} {\bfseries 634}
  (2005) 14--19}, \href{http://arxiv.org/abs/astro-ph/0305189}{{\ttfamily
  arXiv:astro-ph/0305189}}.

\bibitem{Liguori:2005rj}
M.~Liguori, F.~K. Hansen, E.~Komatsu, S.~Matarrese, and A.~Riotto, ``{Testing
  primordial non-gaussianity in cmb anisotropies}'',
  \href{http://dx.doi.org/10.1103/PhysRevD.73.043505}{{\em Phys. Rev. D}
  {\bfseries 73} (2006) 043505},
  \href{http://arxiv.org/abs/astro-ph/0509098}{{\ttfamily
  arXiv:astro-ph/0509098}}.

\bibitem{Shiraishi:2012bh}
M.~Shiraishi, \href{http://dx.doi.org/10.1007/978-4-431-54180-6}{{\em {Probing
  the Early Universe with the CMB Scalar, Vector and Tensor Bispectrum}}}.
\newblock Springer Theses. Springer, 2013.
\newblock \href{http://arxiv.org/abs/1210.2518}{{\ttfamily arXiv:1210.2518
  [astro-ph.CO]}}.

\bibitem{Hu:1997hp}
W.~Hu and M.~J. White, ``{CMB anisotropies: Total angular momentum method}'',
  \href{http://dx.doi.org/10.1103/PhysRevD.56.596}{{\em Phys. Rev. D}
  {\bfseries 56} (1997) 596--615},
  \href{http://arxiv.org/abs/astro-ph/9702170}{{\ttfamily
  arXiv:astro-ph/9702170}}.

\end{thebibliography}\endgroup

\end{document}